\documentclass[a4paper,11pt]{article}
\pdfoutput=1
\usepackage{jheppub}
\usepackage{graphicx}
\usepackage{url}
\usepackage{subcaption}
\usepackage[usenames,dvipsnames]{xcolor}
\usepackage{amsmath, amsthm, mathrsfs, amsfonts, dsfont}
\usepackage{easy-todo}

\newcommand{\figureref}[1]{Fig.~\ref{#1}}
\newcommand{\equationref}[1]{Eq.~\eqref{#1}}
\newcommand{\sectionref}[1]{Sec.~\ref{#1}}
\newcommand{\subsectionref}[1]{Subsection~\ref{#1}}
\newcommand{\appendixref}[1]{Appendix~\ref{#1}}

\def\Caravel{{\textsc{Caravel}}} 
 \renewcommand{\imath}{\mathrm{i}}
\DeclareMathOperator\arctanh{arctanh}
\DeclareMathOperator\arcosh{arcosh}

\makeatletter \newsavebox{\@brx}
\newcommand{\llangle}[1][]{\savebox{\@brx}{\(\m@th{#1\langle}\)}%
  \mathopen{\copy\@brx\mkern2mu\kern-0.9\wd\@brx\usebox{\@brx}}}
\newcommand{\rrangle}[1][]{\savebox{\@brx}{\(\m@th{#1\rangle}\)}%
  \mathclose{\copy\@brx\mkern2mu\kern-0.9\wd\@brx\usebox{\@brx}}}
\makeatother

\title{Spinning Black Hole Scattering at $\mathcal{O}(G^3 S^2)$:
  Casimir Terms, Radial Action and Hidden Symmetry}

\author[1]{Dogan Akpinar,} \author[2]{Fernando Febres Cordero,}
\author[3]{Manfred Kraus,} \author[4]{Michael S.\ Ruf,} \author[1]{and
  Mao Zeng}

\affiliation[1]{Higgs Centre for Theoretical Physics, University of
  Edinburgh, Edinburgh, EH9 3FD, UK} \affiliation[2]{Physics
  Department, Florida State University, Tallahassee, FL 32306-4350,
  USA} \affiliation[3]{Departamento de F\'{i}sica Te\'{o}rica,
  Instituto de F\'{i}sica, \\ Universidad Nacional Aut\'{o}noma de
  M\'{e}xico, Cd. de M\'{e}xico C.P. 04510, M\'{e}xico}
\affiliation[4]{Mani L. Bhaumik Institute for Theoretical Physics,
  University of California at Los Angeles, Los Angeles, CA 90095, USA}

\abstract{We resolve subtleties in calculating the post-Minkowskian
  dynamics of binary systems, as a spin expansion, from massive
  scattering amplitudes of fixed finite spin. In particular, the
  apparently ambiguous spin Casimir terms can be fully determined from
  the gradient of the spin-diagonal part of the amplitudes with
  respect to $S^2 = -s(s+1)\hbar^2$, using an interpolation between
  massive amplitudes with different spin representations. From
  two-loop amplitudes of spin-0 and spin-1 particles minimally coupled
  to gravity, we extract the spin Casimir terms in the conservative scattering
  angle between a spinless and a spinning black hole at
  $\mathcal{O}(G^3 S^2)$, finding agreement with known results in the
  literature. This completes an earlier study [Phys.\ Rev.\ Lett.\ 130
  (2023), 021601] that calculated the non-Casimir terms from
  amplitudes. We also illustrate our methods using a model of spinning
  bodies in electrodynamics, finding agreement between scattering
  amplitude predictions and classical predictions in a \textit{root-Kerr}
  electromagnetic background up to $\mathcal{O}(\alpha^3 S^2)$.  For
  both gravity and electrodynamics, the finite part of the amplitude
  coincides with the two-body radial action in the aligned spin limit,
  generalizing the amplitude-action relation beyond the spinless case.
  Surprisingly, the two-loop amplitude displays a hidden spin-shift
  symmetry in the probe limit, which was previously observed at one
  loop. We conjecture that the symmetry holds to all orders in the
  coupling constant and is a consequence of integrability of Kerr
  orbits in the probe limit at the first few orders in spin.}

\begin{document}
\maketitle
\section{Introduction}
Since the historic detection~\cite{LIGOScientific:2016aoc, LIGOScientific:2017vwq} 
of gravitational waves by the LIGO/VIRGO collaborations, gravitational
wave science has progressed at a rapid pace, with a large number of
black hole and neutron star merger events detected, projected to be
over 200 at the end of the current LIGO O4b run. Next-generation
detectors \cite{Punturo:2010zz, LISA:2017pwj, Reitze:2019iox} will
offer significantly higher sensitivity and require a considerable increase
in precision of theoretical waveform templates~\cite{Purrer:2019jcp}. While the dynamics of spinless binary systems
has enjoyed intense theoretical studies, the effects of spin will be
important for modeling real astrophysical binary systems.

An unexpected new source of theoretical progress for gravitational wave physics
arose from classical limits of scattering amplitudes in quantum field
theory (QFT). This has proven especially powerful for the
\emph{post-Minkowskian expansion}, where one perturbatively expands
in the gravitational constant while keeping full dependence on the velocity
parameter, akin to relativistic amplitude calculations in QFT. Over
the past years, scattering amplitude methods have dramatically
advanced the state of the art for conservative dynamics of spinless
binaries \cite{Cheung:2018wkq, Kosower:2018adc, Bern:2019nnu,
  Bern:2019crd, Cristofoli:2019neg, Bjerrum-Bohr:2019kec,
  Brandhuber:2021eyq, Bern:2021dqo, Bern:2021yeh,
  Damgaard:2023ttc}. See also the reviews \cite{Buonanno:2022pgc, Bjerrum-Bohr:2022blt}. Worldline methods, initially formulated for the
\textit{post-Newtonian expansion} \cite{Goldberger:2004jt}, were also
re-formulated to target the post-Minkowskian expansion and achieved
spectacular results, e.g.\ in Refs.\ \cite{Kalin:2020mvi,
  Kalin:2020fhe, Dlapa:2023hsl, Dlapa:2021npj, Jakobsen:2023oow,
  Mogull:2020sak, Jakobsen:2021smu, Driesse:2024xad}.

Spin effects in the post-Minkowskian expansion have also been studied
through both scattering amplitude and worldline methods in QFT, e.g.\
in Refs.\ \cite{Bini:2017xzy, Bini:2018ywr, Vines:2017hyw,
  Vines:2018gqi, Guevara:2017csg, Guevara:2018wpp, Chung:2018kqs,
  Arkani-Hamed:2019ymq, Guevara:2019fsj, Chung:2019duq,
  Damgaard:2019lfh, Aoude:2020onz, Chung:2020rrz, Guevara:2020xjx,
  Bern:2020buy, Kosmopoulos:2021zoq, Chen:2021qkk,
  FebresCordero:2022jts, Bern:2022kto, Menezes:2022tcs, Riva:2022fru,
  Damgaard:2022jem, Aoude:2022thd, Bautista:2022wjf, Gonzo:2023goe, Aoude:2023vdk,
  Lindwasser:2023zwo, Brandhuber:2023hhl, DeAngelis:2023lvf, Aoude:2023dui, Bohnenblust:2023qmy,
  Jakobsen:2021lvp, Jakobsen:2021zvh, Jakobsen:2022fcj,
  Jakobsen:2022zsx, Jakobsen:2023ndj, Jakobsen:2023hig,
  Heissenberg:2023uvo, Lindwasser:2023dcv, Bautista:2023sdf, Cangemi:2023ysz,
  Brandhuber:2024bnz, Chen:2024mmm, Bhattacharyya:2024kxj, Alaverdian:2024spu}. As a
classical object holds a spin much larger than the quantized unit
$\hbar$, it is natural to consider scattering amplitudes for massive
particles with an arbitrary spin, and carry out the calculation with
analytic dependence on the spin value. In a breakthrough,
Ref.~\cite{Arkani-Hamed:2017jhn} constructed three- and four-point
amplitudes involving massive particles with arbitrary spin dependence
in a \textit{spin-exponentiated} form in the massive spinor-helicity
formalism. Meanwhile, the arbitrary-spin Lagrangian formalism of
Ref.~\cite{Bern:2020buy} writes down the Lagrangian and Feynman
vertices in terms of spin operators acting on any massive spin
representation. These lines of inquiry have led to great success for
calculating binary dynamics at tree and one-loop level, i.e.\
$\mathcal O(G)$ and $\mathcal O(G^2)$. In particular, the one-loop
calculations have reached the 5th order \cite{Bern:2022kto} and even higher/infinite orders \cite{Aoude:2022trd, Aoude:2022thd, Aoude:2023vdk} in spin.

Applying the arbitrary-spin formalisms beyond one loop, however,
remains technically very challenging.  A way to make progress is by
using scattering amplitudes involving particles with fixed spin, which
closely mirrors conventional perturbative QFT calculations for
collider physics and allows straightforward reuse of loop integration
methods developed in the latter context.  It has been well established
that amplitudes involving spin-$s$ matter fields can be used to deduce
spin effects in classical two-body interactions up to the $2s$ order
in the spin expansion (see for example the early work on computing the
post-Newtonian expansion from scattering amplitudes
\cite{Ross:2007zza, Holstein:2008sw, Vaidya:2014kza}). However, an
initially overlooked problem is that the spin Casimir terms are
apparently ambiguous. This can be seen through a quick power-counting
exercise for the classical limit. The exchanged momentum $q$ of each
graviton in the scattering between two massive bodies is the
Fourier-conjugate to the impact parameter $b$, and therefore behaves
as $\mathcal O(\hbar / b)$ in the classical limit, or simply
$\mathcal O(\hbar)$ if only the $\hbar \to 0$ power counting is
concerned. At the same time, the spin quantum number $s$ is no longer
a small (half-) integer and becomes large as $\mathcal O(1/\hbar)$ in
the correspondence limit. Therefore, spin Casimir corrections to
spin-independent classical scattering carry a relative factor of
$q^2 S^2$ which has $\mathcal O(\hbar^0)$ power counting. If one
calculates with massive matter fields of finite spin, such as spin-1
Proca fields \cite{Proca:1936fbw}, Rarita-Schwinger spin-3/2 fermions
\cite{Rarita:1941mf} or Pauli-Fierz spin-2 fields \cite{Fierz:1939ix},
the $q^2 S^2$ terms become indistinguishable from $q^2$ terms which
appear as $\mathcal{O}(\hbar^2)$ quantum corrections of the amplitude.

In this paper, we present the \emph{spin interpolation} method to
solve this problem. By expanding the amplitude beyond the classical
order, we calculate the apparently quantum terms for matter fields
with different spin and extract the \emph{gradient} of the
spin-diagonal part with respect to $S^2 = -s(s+1)$ with $\hbar = 1$ in
the mostly-minus metric convention. This isolates the
spin Casimir terms from the truly quantum terms. The method is general,
though the explicit example in this paper will involve binary systems
with one spinning (Kerr) black hole and one spinless
(Schwarzschild) black hole, and we will calculate the
conservative dynamics up to $\mathcal{O}(G^3)$ and quadratic order in spin. In
the scattering amplitude calculation, the spinless black hole will
be represented as a spin-0 field minimally
coupled to gravity, while for the spinning black hole we will employ 
either a massive spin-1 field or a massive spin-0 field, both
minimally coupled, to obtain the necessary gradients with respect to spin. The spin non-diagonal part of the spin-1
scattering amplitude was used in the earlier work Ref.~\cite{FebresCordero:2022jts} to
calculate the non-Casimir spin structures at $\mathcal{O}(G^3)$, while the
gradient of the spin-diagonal part of the amplitude, obtained from
interpolating between $S^2 = -0\cdot 1=0$ in the spin-0 case and
$S^2=-1\cdot 2=-2$ in the spin-1 case, will be used in the present paper
to extract the spin Casimir term. The validity of our spin
interpolation method will be established by comparing with known
results in the literature~\cite{Jakobsen:2022fcj} for the gravitational case up to $\mathcal{O}(G^3)$,
and by comparing with solutions to classical equations of motion for an analogous
electromagnetic problem of a charged probe particle moving in the
\textit{root-Kerr} background field \cite{Monteiro:2014cda,
  Arkani-Hamed:2019ymq}.

Extracting finite classical quantities from scattering amplitudes
requires careful study. The situation has analogs with collider
physics, where scattering amplitudes have infrared divergences which
only cancel for suitably defined observables that take into account
the finite resolutions of detectors, and, possibly, proper definitions of
the initial states. For applications to classical two-body dynamics,
scattering amplitudes have both infrared divergences and the so-called
\textit{superclassical} divergences, the latter of which are divergences
when one naively takes the $\hbar \to 0$ limit. The divergences cancel
for quantities with meaningful classical limits, such as scattering angles
and two-body interaction potentials. For spinning binary
dynamics, some methods for extracting classical quantities include the
spinning generalization of the Kosower-Maybe-O'Connell (KMOC)
formalism \cite{Maybee:2019jus}, the non-relativistic EFT formalism
generalized to include spin \cite{Bern:2020buy,FebresCordero:2022jts}, and spinning eikonal
exponentiation \cite{Bern:2020buy, Cristofoli:2021jas, Luna:2023uwd,
  Gatica:2023iws}.  A very convenient scheme for calculating the
conservative dynamics of spinless binaries is the amplitude-action
relation \cite{Bern:2021dqo, Bern:2021yeh, Bern:2024adl}, which states
that the classical part of the amplitude, obtained from deleting
superclassical master integrals after soft expansion, coincides with
(the Fourier transform of) the two-body radial action.\footnote{Other
  variants of the amplitude-action relation also exist in the
  literature. Ref.~\cite{Brandhuber:2021eyq} deletes a different set
  of soft-expanded integrals and also arrive at the radial action at
  3PM. Refs.~\cite{Kol:2021jjc, Gonzo:2024zxo} formulate the relation
  in terms of phase shifts of the non-perturbative amplitude in a
  spherical harmonics decomposition.} The scattering angle can then be
calculated as the derivative of the radial action with respect to the
orbital angular momentum. We will refer to this scheme of subtracting
superclassical divergences as the \emph{radial-action-like} scheme,
and refer to the finite quantity left as the \emph{finite remainder} of the amplitude.
For binaries with spin, if the spin vectors
of individual bodies are aligned with the orbital angular momentum,
then the two-body motion is planar, that is non-precessing, similar to
the spinless case. Unsurprisingly, we will see that the
amplitude-action remains valid in the aligned-spin case for both
gravity and electrodynamics. Although it
will be interesting to generalize the amplitude-action relation to
generic spin configurations, the aligned-spin limit is sufficient for
the purpose of extracting spin Casimir terms in the two-body dynamics
at the quadratic order in spin,
and therefore we do not repeat the 
calculation using the spinning potential-EFT and KMOC formalisms carried out at two
loops for non-Casimir terms in the earlier study
\cite{FebresCordero:2022jts}.

A crucial question in the scattering amplitude approach to classical observables is how
massive spinning matter fields are coupled to gravity in order to serve
as an effective point-particle description of Kerr black
holes. Various proposals have appeared from both the point of view of
on-shell amplitudes and the point of view of Lagrangians for arbitrary
spin \cite{Cangemi:2022bew, Bjerrum-Bohr:2023jau, Haddad:2023ylx,
  Alessio:2023kgf, Bjerrum-Bohr:2023iey}. An important observation at
one loop, which corresponds to the second post-Minkowskian (2PM) order, i.e.\ $\mathcal O(G^2)$, is that spin
structures in the finite remainder of the amplitude appear in certain
special combinations. In other words, the coefficients of different
spin structures are linearly dependent. This is emphasized in e.g.\
Ref.~\cite{Aoude:2022trd} and is identified as a hidden symmetry of
the amplitude, the \emph{spin shift
  symmetry}, in Ref.~\cite{Bern:2022kto}. See also Ref.~\cite{Chen:2022clh}. This
symmetry is shown to be eventually broken for Kerr black holes by
solving the Teukolsky equation \cite{Bautista:2022wjf, Bautista:2023sdf}, but
only at higher-than-quartic orders in spin at the one-loop (2PM)
order. This is much higher than the order of spin reached in
state-of-the-art two-loop (3PM) results, and therefore this symmetry
would remain highly relevant if it persists at the 3PM order at low
orders in spin. 
We will show that at the two-loop level through the quadratic
order in the spin of one black hole, 
when using the radial-action-like subtraction scheme, this symmetry is
present in the finite remainder of the amplitude in the probe limit.
The probe limit refers to the limit where either the spinning or the spinless black
hole has a mass far smaller than the other object. By analyzing the
domain of validity of the spin shift symmetry in light of existing
results and new results in this work, we conjecture that the symmetry
is a manifestation of integrability.

This article is organized as follows. In \sectionref{sec:ReviewOfMethods} we 
briefly review the various techniques that we use in our calculation. In 
\sectionref{sec:ClassicalSpin} we discuss the quantum interpretation of spin, show 
how one can make the spin information manifest in amplitudes for processes including 
particles with spin, and present our new method of spin interpolation to extract 
the ambiguous spin Casimir from theories with particles of fixed spin.
In \sectionref{sec:ScatteringAmplitudes} we specify the  
Lagrangians for the theories we consider, how tree amplitudes and loop amplitudes
are determined, and finally present the finite remainders of the classical amplitudes. 
This is done for both electrodynamics and gravity. 
Then in \sectionref{sec:Observables} we discuss 
the integrals needed to extract observables, and present the 
observables that we compute in electrodynamics and gravity. Finally, in 
\sectionref{sec:Conclusion} we give our conclusions and future
outlook, and in particular lay out conjectures regarding the spin
shift symmetry.


\section{Review of methods}
\label{sec:ReviewOfMethods}
In this section we review the main techniques that are employed
in our calculations. We start with choosing a kinematic setup for scattering 
amplitudes in \subsectionref{sec:KinematicSetUp}, followed by 
\subsectionref{sec:ClassicalLim} which gives a brief discussion on extracting the classical 
dynamics from these amplitudes. In \subsectionref{sec:GenUnitarity} 
we review the generalized unitarity method, presenting the analytic
and numeric variants that we use to determine the needed scattering
amplitudes.
Then, \subsectionref{sec:FFDecomposition} explains the form factor 
decomposition used to handle the spinning degrees of freedom in
the scattering amplitudes. 
We also briefly elaborate on the methods used to reduce  
Feynman integrals through integration by parts identities (IBPs) 
in \subsectionref{sec:IBPs}. 
Finally, in \subsectionref{sec:RadialAction}, we review the radial 
action in the context of scalar scattering, which we later extend to 
the case of spinning particles in aligned-spin motion.

\subsection{Kinematic setup}
\label{sec:KinematicSetUp}
\begin{figure}[h!]
 \centering
 \includegraphics[scale=1]{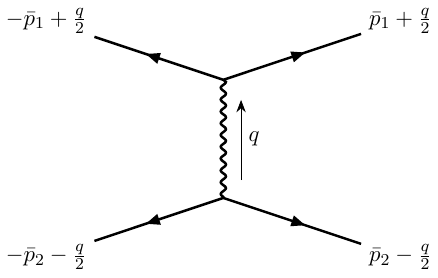}
 \caption{Momentum assignments for the scattering particles. The lower external lines 
 represent massive scalar particles, the upper external 
 lines represent massive vector particles, and wavy 
 lines represent exchanged massless particles. External momenta  
 are taken to be outgoing.}
 \label{fig:kinematics}
\end{figure}

To calculate the dynamics of a binary system with a spinless and spinning 
black hole, we consider the $2 \rightarrow 2$
scattering of a massive vector $V$ of mass $m_1$ and a massive scalar
$\phi$ of mass $m_2$ by exchanging massless particles (gravitons in
Einstein gravity or photons in electrodynamics)
\begin{equation}
  V^\nu(-p_1) + \phi(-p_2) \rightarrow \phi(p_3) + V^\mu(p_4)\;,
  \label{eq:phi_phi_V_V}
\end{equation}
where the signs follow our all-outgoing convention for external momenta.
To facilitate the extraction of classical physics, it is convenient to choose a 
useful parametrization of external variables. Here we choose the external momenta 
as shown in Fig.~\ref{fig:kinematics}, 
which are parametrized by
\begin{equation}
   p_1 = -\bar{p}_1 + \frac{q}{2}\;, \qquad p_2 = -\bar{p}_2 - \frac{q}{2}\;, \qquad p_3 = \bar{p}_2 - \frac{q}{2}\;, 
    \qquad p_4 = \bar{p}_1 + \frac{q}{2}\;. 
    \label{eq:KinematicSetUp}
\end{equation}
The massive vector particles also carry polarization vectors 
$\varepsilon_1^\nu(-p_1)$ and $\varepsilon_4^{\star \mu}(p_4)$, 
where complex conjugation is applied for the polarization vectors of the 
final-state vector particle.

The barred momenta $\bar p_i$ are considered to be much larger than $q$, as explained in
\sectionref{sec:ClassicalLim}, and can be considered as the
average momenta on top of which there are $\mathcal O(q)$
fluctuations. The barred momenta are also orthogonal to $q$,
\begin{equation}
  p_1^2 - p_4^2 = 2 (\bar{p}_1 \cdot q) = 0\;, \qquad p_2^2 - p_3^2 = -2 (\bar{p}_2 \cdot q) = 0\;,
\end{equation}
which follows from on-shell conditions: $p_1^2 = p_4^2 = m_1^2, \  p_2^2 = p_3^2 = m_2^2$. 
Furthermore, we define normalized velocity vectors,
\begin{equation}
 \bar{u}_1^{\mu} = \frac{\bar{p}_1^{\mu}}{\bar{m}_1}\;, \qquad  \bar{u}_2^{\mu} = \frac{\bar{p}_2^{\mu}}{\bar{m}_2}\;,
\end{equation}
with 
\begin{equation}
  \bar{m}_i^2 = \bar{p}_i^2 = m_i^2 - \frac{q^2}{4}\;,
  \label{eq:mBarDef}
\end{equation}
which, by virtue of normalization, satisfy
\begin{equation}
  \bar{u}_1^2 =  \bar{u}_2^2 = 1\;, \qquad  \bar{u}_1 \cdot q =  \bar{u}_2 \cdot q = 0\;.
  \label{eq:velvectorsproperties}
\end{equation}
Finally, we define the kinematic variable
\begin{equation}
   y = \bar{u}_1 \cdot \bar{u}_2 = \frac{\bar{p}_1 \cdot \bar{p}_2} {\bar{m}_1 \bar{m}_2} \;,
   \label{eq:yVarDef}
 \end{equation}
which is a measure of the relative velocity or the
rapidity difference between the two massive particles.

As we will see in \sectionref{sec:ClassicalSpin}, we will need
to perform an interpolation between amplitudes with different spin
representations to obtain spin Casimir terms in the classical two-body
dynamics. Therefore, we will also calculate a scalar-scalar amplitude
that is the counterpart of the scalar-vector amplitude in
Eq.~\eqref{eq:phi_phi_V_V},
\begin{equation}
  \Phi(-p_1) + \phi(-p_2) \rightarrow \phi(p_3) + \Phi(p_4)\;,
  \label{eq:phi_phi_Phi_Phi}
\end{equation}
where $\Phi$ is a scalar particle with mass $m_1$, same as that of the
vector particle in Eq.~\eqref{eq:phi_phi_V_V}.

\subsection{Classical limit}
\label{sec:ClassicalLim}
Since we are interested in classical dynamics, it is appropriate to 
retain the minimal amount of information necessary when describing
scattering processes. To illustrate how this is achieved in practice, let us first 
restrict our discussion to spin-$0$ particles. The distinction between classical and 
quantum information is dictated by an appropriate expansion -- known as the 
\textit{soft} expansion -- 
which follows from the method of regions~\cite{Beneke:1997zp,Smirnov:2012gma}. 
This involves identifying the main scales in the process as
\begin{equation}
\begin{split}
    & \text{Compton length:} \ \ell_c \sim \frac{\hbar}{m}\;, \\
    & \text{Schwarzschild radius:} \ r_s \sim Gm\;, \\
    & \text{Impact parameter:} \ |b| \sim \frac{\hbar}{|q|}\;,
\end{split}
\end{equation}
where $m \sim m_1 + m_2$ is some common mass scale, $q$ is the momentum 
transfer, and $b$ is the relative transverse distance (Fourier conjugate to $q$) 
of the system. Together, these scales define a hierarchy~\cite{Herrmann:2021tct}
\begin{equation}
  \ell_c \ll r_s \ll |b| \quad \Longleftrightarrow \quad \frac{\hbar}{m} \ll Gm \ll \frac{\hbar}{|q|}\;,
\end{equation}
where we keep powers of $\hbar$ for clarity. From there, it is 
straightforward to see
\begin{equation}
    \frac{\ell_c}{|b|} \ll \frac{r_s}{|b|} \ll 1  \quad \Longleftrightarrow \quad \frac{|q|}{m} \ll \frac{Gm|q|}{\hbar} \ll 1\;.
    \label{eq:HierarchyOfScales}
\end{equation}
In the above equation, the smallest quantity $|q|/m$ is identified as
the scaling of $\mathcal O(\hbar)$ quantum corrections. From now on,
we will set $\hbar = 1$ for most of the text, relying on the fact that
a small-$|q|$ expansion can be used as a proxy for the small-$\hbar$
expansion when the mass $m$ is treated as an $\mathcal O(1)$ quantity,
to isolate classical information from quantum corrections.

Within the classical part, the dimensionless perturbative expansion
parameter is $r_s/|b| \sim Gm|q|$, that is the second smallest quantity
in Eq.~\eqref{eq:HierarchyOfScales}. 
Therefore, for each additional order of $G$, we need to expand the amplitude up to one 
additional power of $|q|$. More generally, at a given $\mathcal{O}(G^n)$ we expand the 
scattering amplitude up to $\mathcal{O}(|q|^{n-3})$, in order to extract 
the classical information. An important point is that at one loop and beyond the 
classical order is not simply the leading piece in the expansion. In fact, there are 
superclassical orders -- classically divergent terms -- which start at 
$\mathcal{O}(1/|q|^2)$ and, therefore, precede the classical 
order. These superclassical terms cancel or vanish in observables that have a 
meaningful classical limit, and therefore play no physical role.

When considering spinning external particles, this narrative needs to be extended.
This follows 
from the spin vector associated to the spinning particle $S$ scaling
as $\mathcal{O}(1/|q|)$. The consequence is that, to extract the full classical 
information at a given $\mathcal{O}(G^n)$, we must expand to higher orders in $|q|$
relative to the leading superclassical order. The order needed 
is determined by the fact that 
for processes involving a massive spin-$s$ particle
we expect to probe classical dynamics up to 
$\mathcal{O}(S^{2s})$ in the spin expansion~\cite{Ross:2007zza, Holstein:2008sw, Vaidya:2014kza}. 
As our focus are processes involving spin-$0$ and spin-$1$ particles, 
in order to extract the $\mathcal{O}(S^2)$ corrections, we need to expand to two orders relative 
to the leading superclassical order:
\begin{equation}
\begin{split}
    & \text{~~~~~~tree level:~~~expand to} \ \mathcal{O}(|q|^0)\;, \\
    & \text{one-loop level:~~~expand to} \ \mathcal{O}(|q|^1)\;, \\
    & \text{two-loop level:~~~expand to} \ \mathcal{O}(|q|^2)\;.
\end{split}
\end{equation}
The same discussion applies to electrodynamics. The only change is that the Schwarzschild 
radius becomes the size of a classical particle, $r_s \sim \alpha/m$, where 
$\alpha = e^2/(4\pi)$ is the electromagnetic coupling which is used as the expansion parameter.

\subsection{Generalized unitarity}
\label{sec:GenUnitarity}

In this section we review some aspects of the generalized unitarity
method~\cite{Bern:1994zx,Bern:1994cg,Bern:1997sc,Britto:2004nc} for
the calculation of multi-loop scattering amplitudes in QFT. 
We employ two variants of the method, one more tuned for 
calculations involving analytic integrands and another geared towards
numerical evaluations of the necessary amplitudes. With the former we determine
QED amplitudes up to two-loops, as well as one-loop amplitudes in gravity.
The latter is employed to compute all scattering amplitudes (in QED and
gravity) up to the two-loop level. Numerical evaluations are employed to
determine analytic expressions in a procedure commonly referred to as
functional reconstruction. The two independent approaches we employ
to compute the scattering amplitudes provide (in case of overlap) 
non-trivial validations of our results.

\subsubsection{Analytic calculations}
\label{sec:AnalyticUnitarity}

Generalized unitarity~\cite{Bern:1994zx,Bern:1994cg,Bern:1997sc,Britto:2004nc} allows for the construction of 
loop-level integrands as products of tree-level amplitudes. This follows from
integrands being rational functions, and thereby having simple poles associated
to on-shell propagators. Propagators are put on-shell by replacing the inverse
propagators with a delta function whose argument enforces the on-shell
conditions. Then, we consider the residues of these poles -- or
\textit{generalized cuts} -- to construct integrands as a product of tree-level
amplitudes summed over all physical-states that can cross the 
cut
\begin{equation}
 \mathcal{M}^{(L)}_{\text{residue}} \equiv \sum_{\text{states}} \mathcal{M}^{(0)}_{1}
 \mathcal{M}^{(0)}_{2}\ldots \mathcal{M}^{(0)}_{n-1}\,\mathcal{M}^{(0)}_{n}\;.
 \label{eq:factorization}
\end{equation}
When considering the summation over physical states, one requires transverse projectors for the 
photon and the massive vector boson 
\begin{equation}
	\begin{split}
		& P_{\gamma}^{\mu\nu}(k)\equiv \sum_{\lambda_\gamma} \epsilon_{\lambda_\gamma}^{\star\mu}(k) 
	 	\epsilon_{\lambda_\gamma}^\nu(k) = -\eta^{\mu \nu} + \frac{k^\mu r^\nu + k^\nu r^\mu}{k\cdot r}\;, \\
	 	& P_{\mathrm{V}}^{\mu\nu}(k)\equiv
	 	\sum_{\lambda_{\text{V}}} \varepsilon_{\lambda_{\mathrm{V}}}^{\star\mu}(k)
	 	\varepsilon_{\lambda_{\mathrm{V}}}^{\nu}(k) = - 
	 	\eta^{\mu \nu} + \frac{k^\mu k^\nu}{m_{\text{V}}^2}\;,
	\end{split}
\label{eq:stateprojectors}
\end{equation}
where $\lambda_i$ label physical states, $k$ is the momentum of the cut particle, $r$ is a massless reference momentum, and $m_{\text{V}}$
is the mass of the massive vector boson. Notice that we use $\epsilon_i^\mu$ and $\varepsilon_i^\mu$
to denote the polarization vectors of massless and massive vector
particles, respectively. For cuts involving gravitons we require 
the $D_{\mathrm{s}}$-dimensional graviton transverse projector
\begin{equation}
    P_{h}^{\mu\nu \rho \sigma}(k)\equiv \sum_{\lambda_{h}} \epsilon_{\lambda_{h}}^{* \mu \nu}(k)\epsilon_{\lambda_{h}}^{\rho \sigma}(k) = \frac{1}{2}\left(P_{\gamma}^{\mu \rho}P_{\gamma}^{\nu \sigma} + P_{\gamma}^{\mu \sigma}P_{\gamma}^{\nu \rho}\right) - \frac{1}{D_{\mathrm{s}} - 2}P_{\gamma}^{\mu \nu} P_{\gamma}^{\rho \sigma}\;,
\end{equation}
where $\epsilon_{\lambda_{h}}^{\rho \sigma}(k)$ represent the
polarization tensors for the cut gravitons. 
We have shown the generic $D_s$ dependence in the above equation, but
we specialize to $D_s=d=4-2\epsilon$, i.e.\ conventional dimensional
regularization, in our calculations.
If tree amplitudes in Eq.~\eqref{eq:factorization} satisfy generalized Ward identities then the 
second term in $P_{\gamma}^{\mu\nu}$ can be dropped~\cite{Kosmopoulos:2020pcd}, 
that is we can replace $P_{\gamma}^{\mu \nu}$ by $-\eta^{\mu \nu}$.

We then express the loop integrand
as a linear combination of Feynman
diagrams, each with a set of propagators and a numerator.
The complete quantum integrand can be obtained by a spanning set of cuts,
namely, the set of cuts that allows one to fix every term in the linear
combinations.
Since we are interested in only the classical 
information in the
four-point scattering amplitudes, the number of diagrams that we actually need to consider is
drastically reduced. In particular, we expect no 
contribution from diagrams that have pinched photon/graviton lines, since these 
correspond to zero-point quantum interactions with no classical consequence.
This in turn reduces the number of spanning cuts needed to construct the classically relevant
part of the loop integrand.

\begin{figure}[h!]
  \centering
  \begin{subfigure}{0.32\textwidth}
    \centering
    \includegraphics[scale=1.75]{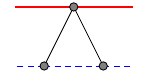}
    \caption{Triangle cut}
    \label{fig:TriangleCut}
  \end{subfigure}
  \begin{subfigure}{0.32\textwidth}
    \centering
    \includegraphics[scale=1.75]{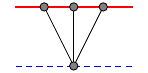}
    \caption{Double triangle cut}
    \label{fig:double_triangle_cut}
  \end{subfigure}
  \begin{subfigure}{0.32\textwidth}
    \centering
    \includegraphics[scale=1.75]{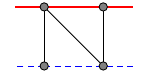}
    \caption{N cut}
    \label{fig:n_cut}
   \end{subfigure}
  \caption{Spanning set of unitarity cuts relevant for conservative scattering at one  (\figureref{fig:TriangleCut}) and two loops (\figureref{fig:double_triangle_cut}, \figureref{fig:n_cut}) for QED. 
      Only topologically inequivalent diagrams are shown.
   	  Thin black lines represent massless particles, while 
   	  thick dashed blue lines and thick red solid lines represent massive particles with different masses. 
      Moreover, all internal lines are cut, and blobs depict appropriate tree 
   	  amplitudes.}
  \label{fig:CutConfigurations}
\end{figure}

At one loop, it is well known that 
the superclassical and classical contributions are the box, crossed box, triangle, 
and inverted triangle diagrams. Here \textit{crossed} refers to the crossing from the 
s-channel to the u-channel, e.g.\ by exchanging $p_1$ and $p_4$.
Triangle (see \figureref{fig:TriangleCut})
and inverted triangle cuts can then be used to 
fix the box, crossed box, and their respective 
triangle integrands.
The two-loop integrand, on the other hand, is more involved as there are many more
diagrams in the linear combination 
(see for example Ref.~\cite{Bern:2019crd}
for details). At this loop level, the relevant cuts are 
the double triangle, inverted double triangle, N, 
horizontally flipped N, and crossed N cuts (see 
\figureref{fig:double_triangle_cut} and \figureref{fig:n_cut}), using a naming
convention that describes what the massless part of the diagram looks like.
Then, the classical two-loop integrand is determined by putting the information
of all cuts together, while excluding any duplicate copies of integrand terms
that are fixed by more than one spanning cut.

\subsubsection{Numerical calculations}
\label{sec:NumericalUnitarity}
We perform numerical calculations of the necessary scattering 
amplitudes employing the multi-loop numerical unitarity 
method~\cite{Ita:2015tya,Abreu:2017idw,Abreu:2017xsl,Abreu:2017hqn}. 
These numerical evaluations, carried out with finite field arithmetic, 
are used to reconstruct analytic expressions (see 
e.g. Refs.~\cite{vonManteuffel:2014ixa,Peraro:2016wsq,Abreu:2018zmy}).
The power of this approach for the study of scattering amplitudes
in gravity has already been demonstrated in the computation of several 
amplitudes for processes involving four and five 
particles~\cite{Abreu:2020lyk,FebresCordero:2022jts,Bohnenblust:2023qmy}.

\begin{figure}
  \centering
  \begin{subfigure}[b]{0.19\textwidth}
  \centering
      \includegraphics{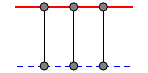}
      \caption{}
      \label{subfig:Cut_2L_7_1}
  \end{subfigure}
  \begin{subfigure}[b]{0.19\textwidth}
  \centering
      \includegraphics{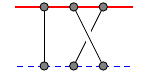}
      \caption{}
      \label{subfig:Cut_2L_7_2}
  \end{subfigure}
  \begin{subfigure}[b]{0.19\textwidth}
  \centering
      \includegraphics{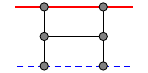}
      \caption{}
      \label{subfig:Cut_2L_7_3}
  \end{subfigure}
  \begin{subfigure}[b]{0.19\textwidth}
  \centering
      \includegraphics{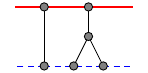}
      \caption{}
      \label{subfig:Cut_2L_7_4}
  \end{subfigure}
  \begin{subfigure}[b]{0.19\textwidth}
  \centering
      \includegraphics{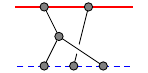}
      \caption{}
      \label{subfig:Cut_2L_7_5}
  \end{subfigure}\\
  \begin{subfigure}[b]{0.19\textwidth}
  \centering
      \includegraphics{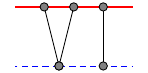}
      \caption{}
      \label{subfig:Cut_2L_6_1}
  \end{subfigure}
  \begin{subfigure}[b]{0.19\textwidth}
  \centering
      \includegraphics{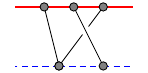}
      \caption{}
      \label{subfig:Cut_2L_6_2}
  \end{subfigure}
  \begin{subfigure}[b]{0.19\textwidth}
  \centering
      \includegraphics{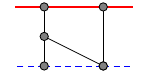}
      \caption{}
      \label{subfig:Cut_2L_6_3}
  \end{subfigure}
  \begin{subfigure}[b]{0.19\textwidth}
  \centering
      \includegraphics{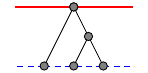}
      \caption{}
      \label{subfig:Cut_2L_6_4}
  \end{subfigure}
  \begin{subfigure}[b]{0.19\textwidth}
  \centering
      \includegraphics{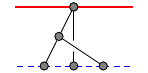}
      \caption{}
      \label{subfig:Cut_2L_6_5}
  \end{subfigure}
  \begin{subfigure}[b]{0.19\textwidth}
  \centering
      \includegraphics{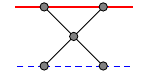}
      \caption{}
      \label{subfig:Cut_2L_6_6}
  \end{subfigure}
  \begin{subfigure}[b]{0.19\textwidth}
  \centering
      \includegraphics{figs/Cut_2L_5_1.pdf}
      \caption{}
      \label{subfig:Cut_2L_5_1}
  \end{subfigure}
  \begin{subfigure}[b]{0.19\textwidth}
  \centering
      \includegraphics{figs/Cut_2L_5_2.pdf}
      \caption{}
      \label{subfig:Cut_2L_5_2}
  \end{subfigure}
  \caption{The propagator structures needed for conservative scattering at the 
two-loop level. The thick solid red and dashed blue lines on the top and bottom of the diagrams 
represent on-shell particles with different masses.  The thin black lines 
represent on-shell massless particles. The vertices of these diagrams specify 
the tree-level amplitudes on the left-hand side of the cut equation~\eqref{eq:cut_equation}. 
Only topologically-inequivalent diagrams are shown. For the QED calculations 
only diagrams \subref{subfig:Cut_2L_7_1}, \subref{subfig:Cut_2L_7_2},
\subref{subfig:Cut_2L_6_1}, \subref{subfig:Cut_2L_6_2}, \subref{subfig:Cut_2L_5_1} 
and \subref{subfig:Cut_2L_5_2} contribute.}    
  \label{fig:twoloop_cut_hierarchy}
\end{figure}

We start with a parametrization of the integrand of the
$L$-loop amplitude $\mathcal{M}^{(L)}(\ell_l)$ (we use $\ell_l$
as shorthand notation for the collection of all loop momenta)
\begin{equation}
 \mathcal{M}^{(L)}(\ell_l) = \sum_{\Gamma} \sum_{i\in Q_\Gamma} 
 c_{\Gamma,i} 
 \frac{m_{\Gamma,i}(\ell_l)}{\prod_{j\in P_\Gamma} \rho_j}\;,
 \label{eq:integrand_param}
\end{equation}
where the outer sum runs over the set of all propagator structures $\Gamma$.
$Q_\Gamma$ is a set of indices for a basis of
functions $m_{\Gamma,i}(\ell_l)$ which parametrizes all possible integrand
numerator insertions in $\Gamma$. Finally, $P_\Gamma$ labels all inverse
propagators $\rho_j$ (which depend implicitly on $\ell_l$) associated with $\Gamma$.
%
The coefficients $c_{\Gamma,i}$ contain all the process- and 
theory-dependent information of the integrand 
$\mathcal{M}^{(L)}(\ell_l)$. To determine them, 
Eq.~\eqref{eq:integrand_param} can be sampled over different values 
of the loop momenta $\ell_l$ to produce a system of linear equations
by which they can be computed.
This procedure simplifies dramatically if one chooses loop-momentum
configurations $\ell_l^\Gamma$ such that the inverse propagators 
$\rho_j$ with $j \in P_\Gamma$ are on-shell, that is
\begin{equation}
 \rho_j(\ell_l^\Gamma) = 0~~\forall~~j\in P_\Gamma\;. 
\end{equation}
Indeed, in the limit $\ell_l\to\ell_l^\Gamma$ the left-hand side of 
Eq.~\eqref{eq:integrand_param} factorizes into a product of tree-level
amplitudes (same as in Eq.~\eqref{eq:factorization}), and from the leading
singularity in this limit one extracts the so-called 
\textit{cut equation}~\cite{Abreu:2017xsl}
\begin{equation}
 \sum_{\text{states}} \mathcal{M}^{(0)}_1(\ell_l^\Gamma) \dots \mathcal{M}^{(0)}_{n_\Gamma}(\ell_l^\Gamma) =
 \sum_{\Gamma^\prime\ge\Gamma,i\in Q_{\Gamma^\prime}}
 \frac{c_{\Gamma^\prime,i}~m_{\Gamma^\prime,i}(\ell_l^\Gamma)}
 {\prod_{j\in (P_{\Gamma^\prime}\setminus P_\Gamma)} \rho_j(\ell_l^\Gamma)}\;,
 \label{eq:cut_equation}
\end{equation}
where $n_\Gamma$ labels the number tree-level amplitudes on the 
left-hand side products (which corresponds to the number of vertices in the 
diagram associated to $\Gamma$), and the sum over the states runs 
over all physical states for the loop particles. On the right-hand side, the sum over $\Gamma^\prime\ge\Gamma$ means a sum over all 
propagator structures $\Gamma^\prime$ that contain all propagators 
in $\Gamma$.
Cut equations can be written hierarchically in such a way that all 
coefficients $c_{\Gamma,i}$ can be efficiently computed even for
field theories with large momentum power counting like gravity.
All cut diagrams needed for the two-loop amplitudes we compute
are shown in Fig.~\ref{fig:twoloop_cut_hierarchy}. For more details
on the numerical unitarity method 
see Refs.~\cite{Ita:2015tya,Abreu:2017idw,Abreu:2017xsl,Abreu:2017hqn}.

In our computations we employ the \Caravel{} framework~\cite{Abreu:2020xvt}, 
which we have extended to handle calculations involving massive particles. 
In the cut equations, the products of tree-level amplitudes, or \textit{cuts}, are computed by
solving off-shell recursion relations in a way analogous to tree-level
Berends-Giele recursion relations~\cite{Berends:1987me}.
The necessary Feynman rules for the amplitudes we consider involve up
to five-point vertices including matter fields and four-graviton 
vertices, see Fig.~\ref{fig:twoloop_cut_hierarchy}. The vertex rules 
are derived from the Lagrangians presented in sections \ref{sec:ScatteringAmplitudesQED} and 
\ref{sec:ScatteringAmplitudesGrav} with the help of the program
\textsc{xAct}~\cite{Brizuela:2008ra,Nutma:2013zea}. 

All numerical calculations need to be performed using finite-dimensional
representations of the matter fields. Full dependence on dimensional
regulators are then restored from multiple evaluations in different
dimensions through a procedure known as dimensional 
reconstruction~\cite{Boughezal:2011br}. More precisely, exploiting 
Lorentz invariance we embed two-loop momentum configurations into a 
6-dimensional space (5-dimensional for one-loop momentum). Then the 
representations of the matter fields are written in integer $D_\mathrm{s}$ 
dimensions, with $D_\mathrm{s} \ge 6$. 
The $D_\mathrm{s}$ dependence implicit on the left-hand side of 
Eq.~\eqref{eq:cut_equation} is then resolved by matching to the ansatz
\begin{equation}
 \sum_{\text{states}} \mathcal{M}^{(0)}_1(\ell_l^\Gamma) \dots \mathcal{M}^{(0)}_{n_\Gamma}(\ell_l^\Gamma)
 = \sum_{n=0}^4 d_{\Gamma,n}(D_\mathrm{s}-2)^{n-3}\;,
\label{eq:dim_rec}
\end{equation}
which we evaluate in multiple integer $D_s$ values to extract the 
$d_{\Gamma,n}$ coefficients. Afterwards we can then set symbolically 
$D_s=4-2\epsilon$.
Note that the source of the poles at $D_\mathrm{s} = 2$ in 
Eq.~\eqref{eq:dim_rec} is the graviton propagator in 
Eq.~\eqref{eq:grav_prop}. 

With all integrand coefficients $c_{\Gamma,i}$ in 
Eq.~\eqref{eq:integrand_param} computed, the amplitude can be 
obtained by performing the integration of the functions in the bases
$Q_\Gamma$ for all propagator structures $\Gamma$. That is, by
integrating 
\begin{equation} 
\int \frac{d^d\ell_1}{(2\pi)^d}\frac{d^d\ell_2}{(2\pi)^d} 
    \frac{m_{\Gamma,i}(\ell_l)}{\prod_{j\in P_\Gamma} \rho_j}\ , 
    \qquad \forall \ i\in Q_\Gamma\ , \qquad \forall \ \Gamma\, .
\end{equation}
For a given $\Gamma$, a powerful choice of basis $Q_\Gamma$ is the 
so-called \textit{master-surface} basis~\cite{Ita:2015tya} (see 
also Ref.~\cite{Abreu:2017hqn}), where these integrations give directly 
zero (the \textit{surface} terms), or master integrals (the 
\textit{master} terms) as determined via integration-by-parts (IBP) 
identities~\cite{Chetyrkin:1981qh}. But in this work we have 
constructed, in an automated fashion within \Caravel{}, bases 
$Q_\Gamma$ built from a maximal set of functions $m_{\Gamma,i}(\ell_l)$ 
which are monomials on irreducible scalar products (ISPs) 
(of the form $\ell_i\cdot p_j$), plus a set of transverse surface terms.
The latter are surface terms that involve loop-momentum degrees of 
freedom which are transverse to the scattering plane. We call these 
bases the \textit{scattering-plane tensor} bases. 
The integrals these bases produce are irreducible under 
Lorentz-invariance relations, but are still IBP reducible. 
Nevertheless, they are particularly suitable as their reduction to
master integrals is  greatly simplified when taking the classical limit 
of the corresponding integrands before IBP reduction. We give more 
details about this procedure in \sectionref{sec:AnalyticReconGrav}.

\subsection{Form factor decomposition}
\label{sec:FFDecomposition}
The $2\to 2$ scattering amplitude for a pair of massive spin-$0$ and 
spin-$1$ particles, Eq.~\eqref{eq:phi_phi_V_V}, can be decomposed into the following 5 linearly
independent tensor structures \cite{FebresCordero:2022jts}
\begin{equation}
\Big\{T_1^{\mu \nu},\ldots,T_5^{\mu \nu}\Big\} = \Big\{ \eta^{\mu \nu},\; q^\mu q^\nu,\;
 q^2 \bar{p}_2^\mu \bar{p}_2^\nu,\; 2\,q_{\phantom{2}}^{[\mu} \bar{p}_2^{\nu]},\;2\,q_{\phantom{2}}^{(\mu} \bar{p}_2^{\nu)}\Big\}\;,
\end{equation}
which form a basis set for the amplitude
\begin{equation}
    \mathcal{M} = \sum_{n=1}^5 M_n T_n\;, \qquad  T_n = \varepsilon^\star_{4\mu} T_n^{\mu \nu} \varepsilon_{1\nu}\;.
    \label{eq:tensorBasis}
  \end{equation}
The convenience of this decomposition is that the entire spin dependence is
encapsulated in the tensors $T_n$ for any theory, while the coefficients $M_n$,
the \textit{form factors},
are theory dependent quantities. We work with parity-invariant theories such as Einstein 
gravity and electrodynamics, and therefore do not include Levi-Civita tensors.
The transversality of the polarization vectors, together
with the momentum parametrization Eq.~\eqref{eq:KinematicSetUp},
allows us to eliminate any $\bar p_1^\mu$ and $\bar p_1^\nu$ from the above tensor 
structures. Moreover, the coefficient of $T_5$, which exists 
in our loop integrand, vanishes upon integration, which is expected as $T_5$ does 
not obey the crossing symmetry of the scattering amplitude.

In the numerical calculations described in
\sectionref{sec:NumericalUnitarity}, we use 5 independent sets of
numerical values for transverse polarization vectors
$(\varepsilon_1^\nu, \varepsilon_4^{\star \mu})$ and numerically invert a
linear system to find the form factors $M_n$ at the integrand level.
For the analytic calculations described in \sectionref{sec:AnalyticUnitarity}, 
we obtain the form factors according to the following procedure.

We multiply the matrix element
with a conjugated tensor and sum over all polarizations
\begin{equation}
  \sum_{\text{spin}}\mathcal{M}T_m^\dagger = \sum_{n=1}^5 M_n \sum_{\text{spin}}T_n T^\dagger_m \equiv \sum_{n=1}^5 M_n \mathcal{P}_{nm}\;,
\end{equation}
where we have defined the projection matrix
\begin{equation}
  \mathcal{P}_{nm} = \sum_{\text{spin}}\left(\varepsilon_{4\mu}^\star T_n^{\mu \nu} \varepsilon_{1\nu}\right) 
  \left(\varepsilon^\star_{1\alpha}T_m^{\dagger\alpha\beta}\varepsilon_{4\beta} \right) 
  = P_{4\mu\beta} T_n^{\mu \nu}P_{1\nu\alpha}T_m^{\dagger \alpha \beta} 
  = \textrm{Tr}\left[P_4 T_n P_1 T_m^\dagger\right]\;,
\end{equation}
with $P_i^{\mu\nu}=P_V^{\mu\nu}(p_i)$, where the latter 
is defined in Eq.~\eqref{eq:stateprojectors}, being the polarization sum for the 
vector particle legs $p_1$ and $p_4$ both with mass $m_1$.
Then,  
\begin{equation}
    \sum_{n=1}^5 M_n \mathcal{P}_{nm} =  \sum_{n=1}^5 M_n P_{4\mu\beta} T_n^{\mu \nu}P_{1\nu\alpha}T_m^{\dagger \alpha \beta} 
    = P_{4\mu\beta} \mathcal{M}^{\mu \nu}P_{1\nu\alpha}T_m^{\dagger \alpha \beta} 
    = \textrm{Tr}\left[P_4 \mathcal{M} P_1 T_m^\dagger\right]\;,
\end{equation}
where the right-hand side defines a vector that corresponds to the form factor
coefficients once acted on with the inverse projection matrix
\begin{equation}
   M_k = \sum_{n,m=1}^5M_n\mathcal{P}_{nm}\left(\mathcal{P}^{-1}\right)_{mk} = \sum_{m=1}^5 \textrm{Tr}\left[P_4 \mathcal{M} P_1 T_m^\dagger\right]\left(\mathcal{P}^{-1}\right)_{mk}\;.
\end{equation}
We have used a $5 \times 5$ projection matrix instead of 
a $4 \times 4$ projection
matrix to explicitly check the vanishing of the coefficient of $T_5$
after integration, rather than making the assumption a priori.
The tensor coefficients are expanded in small 
momentum transfer $|q|$. Notice that the order of $|q|$ to which each form factor $M_n$ needs to be 
expanded is correlated with the particular tensor structure. 

\subsection{Integration-by-parts identities}
\label{sec:IBPs}
A crucial step in acquiring the classically relevant amplitude is reducing the number of 
diagrams to a minimal basis of so-called \textit{master integrals}
(MIs), which can be done using, for example, {\tt
  FIRE6}~\cite{Smirnov:2019qkx, Smirnov:2023yhb}. The result 
is that $L$-loop amplitudes are recast as
\begin{equation}
 i \mathcal{M}^{(L)} = \sum_n c_n \mathcal{I}_n\;,
\end{equation}
where $c_n$ are rational functions of kinematic variables, and the MIs 
have the general form
\begin{equation}
    G_{[\alpha_1, \alpha_2, \ldots,  \alpha_{m-1}, \alpha_m]} = \int \prod_{i = 1}^{L} \left(\mu^{2\epsilon}\frac{d^D \ell_i}{(2\pi)^D}\right) \frac{1}{\rho_1^{\alpha_1} \rho_2^{\alpha_2} \ldots \rho_{m-1}^{\alpha_{m-1}} \rho_m^{\alpha_m}}\;,
    \label{eq:GeneralIBPIntegral}
\end{equation}
with loop momenta $\ell_i$, propagators denoted by $\rho_i$, and integer indices 
$\alpha_i \in \mathbb{Z}$. Loop integrals are regularized using dimensional 
regularization, $D = 4 - 2\epsilon$, which involves an additional factor of 
$\mu^{2\epsilon}$ per loop. 
Then, the values of these MIs can be determined using standard 
techniques in the literature, such as differential 
equations~\cite{Henn:2014qga,Grozin:2015kna,Herrmann:2021tct,Parra-Martinez:2020dzs,Henn:2013pwa,Goncharov:2010jf}. Importantly, 
the MIs are evaluated in the potential region, which is sufficient to capture the 
conservative dynamics.

\subsection{The radial action}
\label{sec:RadialAction}
An essential task in the scattering-amplitude-based program for classical
physics is the extraction of the desired classical observables from the
quantum scattering amplitude.  This is non-trivial as the amplitude itself is
not a classical object, as seen by the fact that even the tree-level amplitude
is divergent as $\hbar\to 0$
\begin{equation}
  \mathcal{M}^{(0)}=\mathcal{O}\left(\frac{1}{\hbar}\right)\,.
\end{equation}
Despite the amplitude diverging in classical processes, the transferred quantum
numbers are, at the same time, generally of the order $\hbar$; for example an
individual radiated classical graviton will carry energy and momentum of the
order $\hbar$.  Typically, this is realized in terms of an exponential
representation of the process, with the exponent matched in perturbation
theory.  The most well-known form of such an exponential representation is the
eikonal, which has gained popularity starting in the late 80's (see
Ref.~\cite{DiVecchia:2023frv} and references therein). More recently, several
related representations have been introduced in the context of the
scattering-amplitude-based program
\cite{Brandhuber:2021kpo,Damgaard:2023ttc,Bern:2021dqo}.

Conceptually it is most straightforward to directly target classical
observables as was put forward by Kosower, Maybee and O'Connell (KMOC) in
Ref.~\cite{Kosower:2018adc}.  The reasoning is simple, yet powerful: suitably
defined quantum observables must have a smooth classical limit. The most
important observable is the so-called \textit{impulse}, which is the change of momentum
during a scattering event between the asymptotic past and future. Then, two
important observables are obtained from projections of the impulse: the
center-of-mass scattering angle\footnote{In general, multiple angles are needed to 
define the scattering event if the scattered particles are not confined 
to a scattering plane, e.g. when spin is involved.}
and the total radiated energy, or, equivalently, the missing energy in the two-body
system of the heavy particles. Similarly, one can compute the total radiated
angular momentum and displacement of the center-of-mass during the scattering
event. In a subsequent paper \cite{Maybee:2019jus} another important
observable in the context of spinning bodies was pointed out. The so-called
spin kick is the change of the orientation (and perhaps magnitude
\cite{Bern:2023ity}) of the spin of the individual particles.  While the KMOC
formalism is conceptually straightforward, in practice it can become cumbersome
due to the fact that different types of contributions have to be assembled and
computed. 

For conservative spinless observables, a powerful method to directly extract
the classical observables from the amplitude, known as the
\emph{amplitude-action relation}, has been put forward in
Ref.~\cite{Bern:2021dqo}.  Subsequently this has proven extremely efficient in
order to extract the scattering angle in a variety of problems up to the fourth
order in perturbation theory
\cite{Bern:2021dqo,Bern:2021yeh,Barack:2023oqp,Bern:2023ccb,Bern:2024adl}.
Remarkably, the relation was also found to hold when leading-order radiative
effects are included \cite{Bern:2021xze}. Later we will show that this
generalizes to problems including spin.

Let us review the
amplitude-action relation for spinless particles \cite{Bern:2021dqo}, which reads
\begin{equation}
  \label{eq:aarelation}
           i\mathcal M(E,\boldsymbol{q}) = \int_J e^{i \boldsymbol b
             \cdot \boldsymbol q} (e^{i I_r(E,J)} - 1) \,,\quad \int_J = |\boldsymbol p| \int d^{D-2} \boldsymbol b\;,
\end{equation}
where $I_r(E,J)$ is the radial action~\cite{Landau:1975pou}, which is defined
through the integral of the radial momentum, $p_r$, along the scattering
trajectory
\begin{equation}
  I_r(E,J)=\int p_r(E,J)\, \mathrm{d}r\;.
\end{equation}
The radial action serves as a generating functional for scattering observables, most notably the scattering angle
\begin{equation}
  \chi(E,J)=-\frac{\partial}{\partial J}I_r(E,J)\,.
  \label{eq:angleradaction}
\end{equation}
Inverting Eq.~\eqref{eq:aarelation}, one obtains the radial action as a
\textit{logarithm of the amplitude}. In perturbation theory it is useful to perform
this matching order by order, starting from abstract expansions in powers of
the coupling constant
\begin{equation}
  \label{eq:pexpansions}
  \mathcal{M}=\sum_{k}\mathcal{M}_k\,,\hskip 1 cm  {\tilde{I}}_r=\sum_{k}{\tilde{I}}_{r,k}\,, \hskip 1 cm \tilde{I}_r = \int_J e^{i \boldsymbol b
             \cdot \boldsymbol q} I_r\,,
\end{equation}
where $\mathcal{M}_k$ and $\tilde{I}_{r,k}$ are of $k$-th order in the coupling constant (in the case
at hand either the fine-structure constant $\alpha$ or Newton's constant $G$).  
The first few orders of Eq.~\eqref{eq:aarelation} read
\begin{align}
  \mathcal{M}_{1}={}&{\tilde{I}}_{r,1}\,,\\
  \label{eq:aarelationPert2}
  \mathcal{M}_{2}={}&{\tilde{I}}_{r,2}+\int_{\boldsymbol{\ell}}\frac{{\tilde{I}}_{r,1}{\tilde{I}}_{r,1}}{Z_1}\,,\\
  \label{eq:aarelationPert3}
  \mathcal{M}_{3}={}&{\tilde{I}}_{r,3}+\int_{\boldsymbol{\ell}}\frac{{\tilde{I}}_{r,2}{\tilde{I}}_{r,1}}{Z_1}+\int_{\boldsymbol{\ell}}\frac{{\tilde{I}}_{r,1}{\tilde{I}}_{r,1}{\tilde{I}}_{r,1}}{Z_1Z_2}\,.
\end{align}
where the exponentiation of the radial action appears as iteration integrals
involving products of lower-order terms of the radial action together with
linearized three-dimensional propagators $Z_i$, the details of which can be
found in aforementioned references for the amplitude-action relation.
Structurally, these relations are similar to the relations one would find, for
example, when matching to an effective field theory
\cite{Neill:2013wsa,Cheung:2018wkq}.  The amplitude is equal to a remainder
$\tilde{I}_r$ plus terms originating from iterations of lower-order objects.
The main difference is that in terms of the amplitude-action relation we have a
direct way to identify the part of the amplitude that would have canceled with
the iteration terms on the right-hand side of Eqs.~\eqref{eq:aarelationPert2}
and \eqref{eq:aarelationPert3}.

To be concrete, when evaluating the necessary Feynman integrals to compute the
amplitude $\mathcal{M}_n$ in perturbation theory, we will generally find
integrals in which, after evaluating all energy integrals, we are left with
integrals that have $Z$-poles.  The radial action in momentum space is then
simply obtained by dropping these terms. For conservative two-loop amplitudes,
the implementation is very simple \cite{Bern:2021xze}. We simply need to drop
the triple iteration master integral named ``III'' in
Ref.~\cite{Parra-Martinez:2020dzs} after soft expansion and IBP reduction, and
the real part of the rest of the amplitude gives the radial action in momentum
space.


\section{Classical spin}
\label{sec:ClassicalSpin}
The $ 2 \rightarrow 2$ amplitudes that we compute include polarization 
vectors corresponding to each external spin-$1$ massive particle. 
Here we discuss how one determines spin-dependent quantities
from these polarized amplitudes. 
\subsection{Extracting spin dependence}
\label{sec:ExtractingSpin}
The first step in making the spin-dependence manifest in amplitudes is to
understand what spin means from a quantum mechanical standpoint. The proposed
quantum mechanical interpretation of the classical spin vector is simply the
expectation value of the Pauli-Lubanski operator~\cite{Maybee:2019jus}
\begin{equation}
   \mathbb{W}_\mu = \frac{1}{2}\epsilon_{\mu \nu \rho \sigma} \mathbb{P}^\nu \mathbb{J}^{\rho \sigma}\;,
\end{equation}
where $\mathbb{P}^\nu$, $\mathbb{J}^{\rho \sigma}$ are the translation and
Lorentz generators, respectively. Explicitly, the spin operator is defined as
\begin{equation}
     \mathbb{S}^\mu \equiv \frac{\mathbb{W}^\mu}{m}\;,
\end{equation}
where $m$ is the mass of the spinning particle. The generators fulfill the Lorentz algebra
\begin{equation}
\begin{split}
  & [\mathbb{J}^{\mu \nu},\mathbb{P}^\rho] = i (\eta^{\mu \rho}\mathbb{P}^\nu - \eta^{\nu \rho}\mathbb{P}^\mu)\;, \\
  & [\mathbb{J}^{\mu \nu},\mathbb{J}^{\rho \sigma}] = i(\eta^{\nu \rho}\mathbb{J}^{\mu \sigma} - \eta^{\mu \rho}\mathbb{J}^{\nu \sigma} - \eta^{\nu \sigma}\mathbb{J}^{\mu \rho} + \eta^{\mu \sigma}\mathbb{J}^{\nu \rho})\;,
\end{split}
\end{equation}
and one finds that 
\begin{equation}
 [\mathbb{P}^\mu, \mathbb{W}^\nu] = 0\;, \qquad 
 [\mathbb{J}^{\mu \nu},\mathbb{W}^\rho] = i (\eta^{\mu \rho}\mathbb{W}^\nu - \eta^{\nu \rho}\mathbb{W}^\mu)\;,
\end{equation}
where the latter is expected since $\mathbb{W}^\mu$ is a vector operator. 
One can further determine that
\begin{equation}
    [\mathbb{W}^\mu, \mathbb{W}^\nu] = -i \epsilon^{\mu \nu \rho \sigma}\mathbb{W}_\rho \mathbb{P}_\sigma\;.
\end{equation}
In the rest-frame of the massive particle we find $\mathbb{W}^0 = 0$, which
implies that the remaining generators satisfy 
\begin{equation}
    [\mathbb{W}^i, \mathbb{W}^j] = -i m \epsilon^{i j k }\mathbb{W}_k\;.
\end{equation}
The important observation is that these operators are precisely the 
generators of the little group, which serves to justify the chosen interpretation.

Now the goal is to define a spin vector in terms of polarization
vectors.  To do this, we consider a massive spin-$1$ particle with
momentum $\bar p_1^{\,\prime \mu} = \bar p_1^{\,\mu}(m_1/\bar m_1)$, which
will be taken as the central value of the momentum in the wave packet
we consider within the KMOC formalism. For readers familiar with
heavy-particle effective theory \cite{Aoude:2020onz, Aoude:2022trd},
the central momentum is analogous to the large label momentum in 
such formalism.
The spin vector, as an operator in spin states $i, j$ of the 
little-group representation, takes the
form~\cite{Maybee:2019jus,Chung:2019duq}
\begin{equation}
    \mathbb S_{ij}^\mu =  -\frac{i}{m_1} \epsilon^{\mu \nu \alpha \beta} \bar p^{\,\prime}_{1\nu} \bar \varepsilon^{\,\star}_{i \alpha}(\bar{p}_1^{\,\prime}) \bar \varepsilon_{j \beta}(\bar{p}_1^{\,\prime})\;,
    \label{eq:SpinVectorOperator}
\end{equation}
where $\bar \varepsilon_{j \beta}$ denotes a polarization vector with little-group index $j$ and Lorentz index $\beta$. Consider a massive particle with polarization vector
\begin{equation}
  \bar \varepsilon^{\,\mu} = \sum_j w_j \bar \varepsilon_j^{\,\mu}\ ,
  \label{eq:PolVecWeightedSum}
\end{equation}
that is a complex weighted sum of polarization vectors of different spin
states, then the expectation value of the spin vector is
\begin{equation}
  S^\mu = -\frac{i}{m_1} \epsilon^{\mu \nu \alpha \beta} \bar p_{1\nu}^{\,\prime} \, w_i^{\,\star} \bar \varepsilon^{\,\star}_{i \alpha}(\bar p_1^{\,\prime}) \, w_j \bar \varepsilon_{j \beta}(\bar p_1^{\,\prime})
  = -\frac{i}{m_1} \epsilon^{\mu \nu \alpha \beta} \bar p^{\,\prime}_{1\nu} \bar \varepsilon^{\,\star}_\alpha(\bar p_1^{\,\prime}) \bar \varepsilon_\beta(\bar p_1^{\,\prime})\;.
    \label{eq:SpinVectorPolVecs}
\end{equation}
We refer the
reader to Appendix B of Ref.~\cite{Maybee:2019jus} for more details on the
derivation. Given that the wave packet for an incoming massive particle within the KMOC formalism has a spread in momentum,
the polarization vectors $\varepsilon_1(-p_1)$ and $\varepsilon^\star_4(p_4)$ in the wave packet
can be defined by Lorentz transformations of $\bar \varepsilon(\bar p_1^{\,\prime})$. 
The Lorentz transformations take the central value of the momentum in the wave packet, $\bar p_1^{\,\prime}$,
to $-p_1 = \bar p_1 - q/2$ and $p_4 = \bar p_1 + q/2$, respectively,
and trivially preserve the transversality condition of the polarization vectors\footnote{Note that this is
different to \cite{Maybee:2019jus} where they boost $p_4$ to $p_1$, as we treat the two momenta in a symmetric manner.}.
The explicit derivation of the transformation is shown in \appendixref{appendix:CovariantBoostedpolarizations}. 
In this way, the transformed polarization vectors take the form
\begin{equation}
\begin{split}
  & \varepsilon^\mu_{1} =\Lambda(\psi)^{\mu}_{~\nu}\bar{\varepsilon}^\nu = \bar{\varepsilon}^\mu + A(q\cdot\bar{\varepsilon}) \bar{p}_1^\mu + B (q\cdot\bar{\varepsilon}) q^\mu\;, \\
  & \varepsilon^{\star\mu}_{4} =\Lambda(-\psi)^{\mu}_{~\nu}\bar{\varepsilon}^{\,\star\nu} = \bar{\varepsilon}^{\,\star\mu} - A(q\cdot\bar{\varepsilon}^{\,\star}) \bar{p}_1^\mu + B(q\cdot\bar{\varepsilon}^{\,\star}) q^\mu\;,
\end{split}
\label{eq:polarizationBoost}
\end{equation}
where the boost coefficients are 
\begin{equation}
 A = \frac{1}{\sqrt{-q^2\bar{m}_1^2}}\sinh(\psi)\;, \qquad\qquad  B = \frac{1}{q^2}\left(\cosh(\psi) - 1\right)\;, 
\end{equation}
with
\begin{equation}
 \psi = \arctanh\left(\frac{\sqrt{-q^2}}{{2\bar{m}_1}}\right)\;.
\end{equation}
From the spin vector as an operator in spin states of a definite momentum, Eq.~\eqref{eq:SpinVectorOperator}, the squared spin operator follows as 
\begin{equation}
	\begin{split}
    (\mathbb S^\mu \mathbb S^\nu)_{i j} & = -\frac{1}{m_1^2} \epsilon^{\mu \alpha \beta \gamma} \epsilon^{\nu \lambda \rho \sigma} \bar{p}^{\,\prime}_{1\alpha}\bar{p}^{\,\prime}_{1\lambda}\bar{\varepsilon}^{\,\star}_{i \beta}\bar{\varepsilon}_{j \sigma}\sum_{k} \bar{\varepsilon}^{\,\star}_{k \gamma}\bar{\varepsilon}_{k \rho} \\
    & = -\bar{\varepsilon}^{\,\star\mu}_{i}\bar{\varepsilon}^\nu_{j} - \delta_{ij}\left(\eta^{\mu \nu} - \frac{\bar{p}_1^{\,\prime\mu} \bar{p}_1^{\,\prime\nu}}{m_1^2} \right)\;.
    \end{split}
\end{equation}
These equations can be used to define the linear-in-spin and quadratic-in-spin term for a general weighted sum of individual spin states, Eq.~\eqref{eq:PolVecWeightedSum}, following the same manipulations of Eq.~\eqref{eq:SpinVectorPolVecs} and using the normalization $\sum_i w_i^\star w_i = 1$,
\begin{equation}
\begin{split}
   & \bar{\varepsilon}^{\,\star\mu} (\bar p_1^{\,\prime}) \bar{\varepsilon}^\nu (\bar p_1^{\,\prime}) - \bar{\varepsilon}^{\,\star\nu} (\bar p_1^{\,\prime}) \bar{\varepsilon}^\mu (\bar p_1^{\,\prime})  = \frac{i}{m_1}\epsilon^{\mu \nu \alpha \beta} \bar{p}_{1 \alpha}^{\,\prime} S_{\beta}\;, \\
   & \bar{\varepsilon}^{\,\star\mu} (\bar p_1^{\,\prime}) \bar{\varepsilon}^\nu (\bar p_1^{\,\prime}) + \bar{\varepsilon}^{\,\star\nu} (\bar p_1^{\,\prime}) \bar{\varepsilon}^\mu (\bar p_1^{\,\prime}) = -(S^\mu S^\nu + S^\nu S^\mu)  - 2 \left(\eta^{\mu \nu} - \frac{\bar{p}_1^{\,\prime\mu} \bar{p}_1^{\,\prime\nu}}{m_1^2}\right)\;.
\end{split}
\end{equation}
If we average over the two expressions we find
\begin{equation}
    \begin{split}
        \bar{\varepsilon}^{\,\star\mu} \bar{\varepsilon}^\nu & = - \left(\eta^{\mu \nu} - \frac{\bar{p}_1^{\,\prime\mu} \bar{p}_1^{\,\prime\nu}} {m_1^2}\right) + \frac{i}{2m_1}\epsilon^{\mu \nu \alpha \beta} \bar{p}_{1 \alpha}^{\,\prime} S_{\beta} - \frac{1}{2}(S^\mu S^\nu + S^\nu S^\mu)\;,
    \end{split}
    \label{eq: Spin1Decomposition}
\end{equation}
where the first term contains information on both the spin-independent and Casimir structures, 
through the identification $\eta_{\mu \nu}(S^{\mu}S^{\nu}) = -2$ for the spin-1 representation~\cite{Chung:2019duq}. 
This is a crucial point, since that allows us to extract the full 
Casimir dependence.

For completeness, let us discuss how one can re-express the tensor structures of 
\subsectionref{sec:FFDecomposition}
in terms of  spin vectors. As explained, the first step is performing 
the appropriate Lorentz transformations, where only the leading 
contributions in $q$ are needed
\begin{equation}
 A = \frac{1}{2\bar{m}_1^2} + \mathcal{O}(q)\;, \qquad B = -\frac{1}{8\bar{m}_1^2} + \mathcal{O}(q)\;.
\end{equation}
Then \equationref{eq: Spin1Decomposition} recasts the tensor structures of 
\equationref{eq:tensorBasis} as 
\begin{equation}
	\begin{split}
		& T_1 = -1 + \frac{q^2}{2\bar{m}_1^2} + \frac{1}{2\bar{m}_1^2}(q \cdot S)^2\;, \\
		& T_2 = -q^2 - (q \cdot S)^2\;, \\[8pt]
		& T_3 = \bar{m}_2^2(y^2-1)q^2 - \bar{m}_2^2q^2(\bar{u}_2 \cdot S)^2\;, \\[4pt]
		& T_4 = -i \bar{m}_2 \epsilon^{\mu \nu \rho \sigma} q_\mu \bar{u}_{1\nu}\bar{u}_{2\rho} S_\sigma +\frac{y\bar{m}_2}{\bar{m}_1}q^2 + \frac{y\bar{m}_2}{\bar{m}_1}(q \cdot S)^2\;,
	\end{split}
\end{equation}
which has been evaluated in $D=4$ dimensions. 
Once again, we emphasize that the terms with no spin dependence
contain information about both spin-independent and spin Casimir structures.
Another important point is that the coefficient of $T_1 =
\varepsilon_4^\star \cdot \varepsilon_1$ in the leading small-$|q|$ limit is 
expected to be the negative of the spin-$0$ amplitude, since it corresponds to the 
spin-independent piece of the spin-$1$ amplitude but picks up a minus
sign from the mostly-minus metric signature.

\subsection{Spin interpolation}
\label{sec:SpinInterpolation}

When extracting the spin dependence of classical dynamics, fixed-spin
theories have an apparent ambiguity that stems from the square of
the spin vector $S^2= -s(s+1)$. The consequence is
that any explicit spin Casimir terms are simply numbers, like $S^2 = -2$ for
$s = 1$, and so it seems that we are missing the necessary information
to faithfully capture the spin dependence. In fact, the combination
$q^2 S^2$ is what actually appears, as this combination is of
$\mathcal O(\hbar^0)$ in the classical limit where the spin quantum
number becomes large. This becomes proportional to $q^2$ for a fixed
low spin and is seemingly indistinguishable from the quantum
corrections in the amplitude.
This is one motivation to turn to generic-spin theories
or worldline EFTs, which make the spin structures manifest from the very
start, thereby bypassing the problem altogether.

Here we propose a new method that allows one to extract the full spin
dependence starting from fixed-spin theories. There are two main observations
that allow this. The first is that spin Casimir terms can, naively, be hidden 
as quantum suppressed terms in the spin-independent structure. The second is that 
there exists a form of \textit{spin universality} which makes lower-spin amplitudes
re-appear in certain terms of higher-spin amplitudes. 
To explain how this method works we will focus on the scattering
between a spinless black hole and a spinning black hole, up to
quadratic orders in spin.

At tree-level the $q^2 S^2$ term cancels the $1/q^2$ pole of the tree
amplitude and leaves no non-analytic dependence on $q$ required for classical interactions, so
the spin Casimir ambiguity only plays a role at one-loop order and beyond.
Schematically, after suitable subtraction of superclassical
divergences (the details of which do not matter), the finite remainder of the four-point amplitudes for
the scattering of a scalar particle and a spinning particle is
captured by the following ansatz which holds for \emph{any} spin representation,
\begin{equation}
 \mathcal{M}  = \sum_{n, i} c^{(n,i)} Q^{(n,i)}\;, 
 \label{eq: GeneralAnsatz}
\end{equation}
where $n$ is the spin power in a given structure, and $i$ labels all spin structures 
with a given $n$. The coefficients $c^{(n,i)}$ are generally series expansions in the coupling constant.
Expanding Eq.~\eqref{eq: GeneralAnsatz} in
$|q|$, the spin structures are
\begin{align}
\displaybreak[0]
  & Q^{(0,1)} = 1 + c_{\text{quantum}} q^2 + \mathcal O(q^4)\; , \nonumber \\
  & Q^{(1,1)} = i\epsilon^{\mu \nu \rho \sigma} q_\mu \bar{u}_{1\nu}\bar{u}_{2\rho} a_\sigma \, [ 1 + \mathcal O(q^2)] \;, \nonumber \\
  & Q^{(2,1)} = (q \cdot a)^2 \, [ 1 + \mathcal O(q^2)] \;, \label{eq: SpinStructures}\\ 
  &Q^{(2,2)} = q^2 (\bar{u}_2 \cdot a)^2 \, [ 1 + \mathcal O(q^2)] \;, \nonumber \\ 
  &Q^{(2,3)} = q^2 a^2 \, [ 1 + \mathcal O(q^2)] \;, \nonumber 
\end{align}
where we defined the normalized spin vector,
$ a^\mu \equiv S^\mu/\bar{m}_1$, to make the mass dependence uniform
across all the spin structures. The spin-diagonal part of the
amplitude, again expanded in $|q|$, is
\begin{equation}
  \label{eq:ampSpinDiagonal}
  \mathcal M_{\text{spin-diagonal}} = c^{(0,1)} (1 +
  c_{\text{quantum}} q^2 ) + c^{(2,3)}
  q^2 a^2 + \mathcal{O}(q^4)\;,
\end{equation}
which contains a spinless term, a quantum correction term, and a
spin Casimir term. The spin Casimir term can be isolated from the quantum
correction term at $\mathcal O(|q|^2)$
by an interpolation between the spin-1 amplitude and the spin-0
amplitude\footnote{Recall that we always refer to the spin of the particle with mass $m_1$, while
keeping the $m_2$ particle to be a scalar, as we are only examining
the spin effects of the $m_1$ particle for illustration.}. More explicitly, using
$S^2 = -s(s+1)$, we have
\begin{equation}
  \mathcal
  M_{\text{spin-diagonal}} \Big |_{s=1}  - \mathcal M_{\text{spin-diagonal}}
  \Big |_{s=0} = -2  c^{(2,3)} q^2 / \bar m_1^2\; ,
  \label{eq:spin1MinusSpin0}
\end{equation}
where this identity is understood to $\mathcal{O}(q^4)$.
Therefore, we unambiguously identify the spin Casimir coefficient
$c^{(2,3)}$ from the naively quantum-suppressed $\mathcal O(|q|^2)$ part
of the amplitude, as desired. Here we have relied on spin universality
which ensures that the spinless term and the quantum correction term
are identical for $s=1$ and $s=0$, and therefore cancel out in the
difference on the left-hand side of Eq.~\eqref{eq:spin1MinusSpin0}. To reiterate,
the spin Casimir term is the \emph{gradient} term, and the
quantum correction term is the \emph{non-gradient} term, of the
$\mathcal O(|q|^2)$ corrections to the spin-diagonal part of the amplitude.

As we will see later, with this procedure we faithfully determine the spin
Casimir up to the two-loop level, that is up to the third order in the coupling
constant, for both gravity and electrodynamics.  The method can be
extended to higher spins by performing the same identification for the
other spin structures, to the appropriate order in $|q|$, and then
fixing the coefficients using a tower of amplitudes with different
spins. This provides a systematic method for determining the full spin
dependence of classical two-body dynamics from scattering amplitudes of fixed-spin theories.

\section{Scattering amplitudes}
\label{sec:ScatteringAmplitudes}
In this section we detail the computation of the $2\to 2$ amplitudes that
we need for our studies in electrodynamics and Einstein gravity.
\subsection{Electrodynamics}
\label{sec:ScatteringAmplitudesQED}
The dynamics of charged scalars and vectors, minimally coupled to photons, can be described 
by 
\begin{equation}
  \mathcal{L}_{\mathrm{EM}} = \mathcal{L}_{\text{Maxwell}} + \mathcal{L}_{\text{scalar}} + \mathcal{L}_{\text{vector}}\;.
  \label{eq: EMLagrangian}
\end{equation}
The Maxwell piece is simply the kinetic term for photons
\begin{equation}
  \mathcal{L}_{\text{Maxwell}} = -\frac{1}{4}F_{\mu \nu}F^{\mu \nu}\;, \qquad 
  F_{\mu \nu} = \partial_\mu A_\nu -\partial_\nu A_\mu\;,
\end{equation}
where $A_\mu$ is the photon field. The scalar and 
vector pieces include kinetic terms and the relevant 
interaction terms
\begin{equation}
    \mathcal{L}_{\text{scalar}} = (D_\mu \phi)^{\dagger}(D^\mu \phi) - m_{\phi}^2 \phi^{\dagger}\phi\;, 
\end{equation}
\begin{equation}
\begin{aligned}
      \mathcal{L}_{\text{vector}} = - \frac{1}{2}V^\dagger _{\mu \nu}V^{\mu \nu} + m_{\mbox{\tiny V}}^2 V^\dagger _{\mu} V^{\mu} + i e & V^{\dagger \mu} V_{\mu \nu}A^\nu 
        - i e V^{\dagger }_{\mu \nu}A^\nu V^{\mu} - i e V^{\dagger \mu}F_{\mu \nu}V^{\nu} \\ 
        & + e^2 A_\mu A_\nu V^\mu V^{\dagger \nu} - e^2A^2 V^\mu V^\dagger_\mu\;,
\end{aligned}
\end{equation}
where the massive scalar field $\phi$ and massive vector field $V_\mu$ have masses 
$m_\phi$ and $m_{\mbox{\tiny V}}$ respectively, 
$D_\mu = \partial_\mu + i eA_\mu$ is the usual covariant derivative, and $
V_{\mu \nu}$ is the massive vector field strength
\begin{equation}
	V_{\mu \nu} = \partial_\mu V_\nu - \partial_\nu V_\mu\;.
	\label{eq:MassiveVectorFieldStrength}
\end{equation}
In particular, $\mathcal{L}_{\text{vector}}$ is determined from spontaneous
symmetry breaking a $\mathrm{SU}(2)$ gauge theory, which leads, essentially,
to a theory of $W$ bosons. This differs from the Maxwell-Proca 
Lagrangian by the additional operator 
$-i e V^{\dagger \mu}F_{\mu \nu}V^{\nu}$~\cite{Bautista:2022ewu}, which 
is needed to obtain the expected root-Kerr three-point and 
Compton amplitudes~\cite{Chiodaroli:2021eug,Cangemi:2023ysz}.
\subsubsection{Loop amplitude calculation}
\label{sec:TreesQED}
We start by discussing the relevant tree-level amplitudes needed to compute
the one- and two-loop amplitudes in QED. The
three-point and four-point amplitudes are easily obtained with Feynman
diagrams.  The five-point amplitude in $D$-dimensions is obtained from the
dimensional reduction \cite{Bern:2019crd} of a $(D+2)$-dimensional massless
Yang-Mills amplitudes with an $SO(3)$ gauge group, where the dimensional
reduction accomplishes the spontaneous symmetry breaking and generation of
masses. The massless Yang-Mill amplitudes required are conveniently taken from
the software {\tt IncreasingTrees} developed by Edison and
Teng~\cite{Edison:2020ehu}, which allows one to compute tree-level gluon and
graviton amplitudes to high multiplicity.  The output from {\tt
IncreasingTrees} is organized in terms of the distinct diagrams that contribute
to the five-point Compton amplitude, which facilitates the two-loop amplitude
construction via generalized unitarity.

Dimensional reduction takes a $(D+2)$-dimensional massless theory and 
compactifies 2 dimensions to generate a $D$-dimensional theory involving massive particles. In 
order to generate these massive particles, the polarization vectors and momenta of the external gluons 
need to be chosen accordingly. To illustrate this, let us label the momenta and polarization of 
each particle as $k_i$ and $\epsilon_i$ in the $(D+2)$-dimensional theory. We choose to make particles $1$ 
and $4$ into massive scalars and vectors. To this end, we choose the polarization vectors as  
\begin{equation}
\begin{split}
   \text{scalar}&: \ \epsilon_1^A = \epsilon_4^A = (0,\cdots,0,1)\;, \\
   \text{vector}&: \ \epsilon_1^A = (\varepsilon_1^\mu,0,0), \ \epsilon_4^A = (\varepsilon_4^{\star\mu},0,0)\;,
\end{split}
\end{equation}
and massless momenta as $k_1^A = (p_1^\mu, m, 0), \ k_4^A = (p_4^\mu, -m, 0)$. All other 
polarization vectors and momenta take the form $\epsilon_i^A = (\epsilon_i^\mu,0,0)$,
$k_i^A = (k_i^\mu, 0, 0)$. Here, variables with Latin and Greek indices live in the $(D+2)$- and 
$D$-dimensional theories, respectively. The consequence is that from the perspective of the 
$D$-dimensional theory particles $1$ and $4$ are massive, whereas from the perspective of the 
$(D+2)$-dimensional theory they are massless.

Taking the Abelian limit follows by simply choosing the color indices
of massive and massless particles to be, with the three indices of the
fundamental representation of $SO(3)$,
$a_1 = 1, a_4 =2, a_2 = a_3 = a_5 = 3$. The result is that all color
factors $f^{abc}$ become $-1$, $0$ or $1$, and the resulting amplitude has particles 1 and 4 being
massive $W^\pm$ bosons whereas particles 2, 3 and 5 are photons.

Having calculated tree amplitudes, we use generalized unitarity
as reviewed in \sectionref{sec:AnalyticUnitarity}, to
construct one- and two-loop integrands for the scalar-vector amplitude
Eq.~\eqref{eq:phi_phi_V_V} and scalar-scalar amplitude
Eq.~\eqref{eq:phi_phi_Phi_Phi}. We also have performed numerical evaluations
of all QED integrands with the \Caravel{} framework, as described in 
\sectionref{sec:NumericalUnitarity}, and found agreement with the analytic results. 
Then, we proceed to perform the soft expansion
of the integrand and IBP reduction, before
substituting master integrals by their values in the potential region,
following the procedure in Ref.~\cite{Parra-Martinez:2020dzs}. This includes a
small refinement of fixing the symmetry factors for energy integrals
at the level of three-dimensional master integrals appearing in the
boundary conditions of differential equations \cite{Bern:2021dqo,
  Bern:2023ccb, Bern:2024adl} (see also Ref.~\cite{Cho:2023kux} which
corrects typos in some two-loop integrals of
Ref.~\cite{Parra-Martinez:2020dzs}, albeit using the mostly-plus
metric convention). Then we obtain the finite remainder of the amplitude
using the radial-action-like scheme presented in
\sectionref{sec:RadialAction}, by dropping the triple-iteration master
integral ``III'' and keeping only the real part of the remaining
amplitude. We carry out the spin interpolation procedure, explained in
\sectionref{sec:SpinInterpolation}, and the results are presented
next.
\subsubsection{Classical amplitude and spin-shift symmetry}
\label{sec:AmplitudeQED}
The finite remainder of the four-point tree, one-, and two-loop QED
amplitudes for scalar-vector scattering Eq.~\eqref{eq:phi_phi_V_V} take the
form
\begin{equation}
\begin{split}
 & \mathcal{A}^{(0)}_{\text{finite}} = \frac{\alpha_{\textrm{eff}}}{(-q^2)}  \sum_{n = 1}^{4} A_n^{(0)}T_n + \mathcal{O}(q^0)\;, \\
 & \mathcal{A}^{(L)}_{\text{finite}} = \frac{\alpha_{\textrm{eff}}^{1+L}}{(-q^2)^{1-L/2+L\epsilon}}\sum_{n = 1}^{4}A_n^{(L)} T_n + \text{subleading terms in small } q\;,
\end{split}
\label{eq:ampForm}
\end{equation}
where we define $\alpha_{\textrm{eff}}= q_\phi q_{\mbox{\tiny V}} \alpha$ together 
with $\alpha = e^2/(4\pi)$. Here $A_1^{(L)}$ is separated into two pieces 
\begin{equation}
 A_1^{(L)} = -A_{\text{spin}-0}^{(L)} -q^2 A_{\text{gradient}}^{(L)} + \ldots \;,
 \label{eq:InterpolatedT1Coeff}
\end{equation}
where \ldots \ denote truly quantum-suppressed non-gradient terms.
The first term in Eq.~\eqref{eq:InterpolatedT1Coeff} is the negative of 
the spin-$0$ amplitude, whereas the second term is the gradient
term which is ultimately identified as a spin Casimir term.
The explicit coefficients can be found in the supplemental material file {\tt finiteAmplitudes.m}.

To express the above amplitudes in terms of spin vectors we follow the discussion in 
\sectionref{sec:ClassicalSpin}. The resulting finite remainder,
written in a form for massive particles of any spin, reads
\begin{equation}
 \mathcal{A}_{\text{finite}}^{(L)} = \frac{\alpha_{\textrm{eff}}^{1+L}}{(-q^2)^{1-L/2+L\epsilon}}\sum_{n,i}^{} c^{(L,n,i)} Q^{(n,i)}\;,
 \label{eq:QED_finite}
\end{equation}
where $L$ denotes the loop order and the structures $Q^{(n,i)}$ are defined in 
\equationref{eq: GeneralAnsatz}.
Here we truncate at the quadratic order in spin, $n=2$.
Then, the coefficients are as follows. At tree level we have
\begin{equation}
\begin{split}
 &c^{(0,0,1)} = -16y \pi \bar{m}_1\bar{m}_2\;, \qquad c^{(0,1,1)} = -16 \pi\bar{m}_1\bar{m}_2\;, \\
 & c^{(0,2,1)} = - 8 \pi \bar{m}_1\bar{m}_2\;, \qquad ~~~ c^{(0,2,2)} = c^{(0,2,3)} = 0\;.
 \end{split}
\end{equation}
At one loop we find
\begin{equation}
\begin{split}
 & c^{(1,0,1)} = 4\pi^2(\bar{m}_1 + \bar{m}_2)\;,  \\
 & c^{(1,1,1)} = 4\pi^2y\frac{2\bar{m}_1 + \bar{m}_2}{(y^2-1)}\;,  \\
 & c^{(1,2,1)} = - c^{(1,2,3)} = \pi^2\frac{(5y^2-3)\bar{m}_1+2y^2 \bar{m}_2}{(y^2-1)}\;, \\
 &  c^{(1,2,2)} = - \pi^2\frac{4(2y^2-1)\bar{m}_1+2(y^2+1)\bar{m}_2}{(y^2-1)^2}\;,
\end{split}
\end{equation}
which is the sum of two types of mass structures: $\bar m_1$ which dominates in the spinless probe limit, 
$\bar{m}_1 \gg \bar{m}_2$; and $\bar m_2$ which dominates in the spinning probe limit, $\bar{m}_2 \gg \bar{m}_1$.
In other words, the sum of the two opposite probe limits completely fixes the one-loop amplitude. Practically, one can extract each probe limit from the amplitude by simply taking the limit as the heavy mass goes to infinity.
The two-loop result, on the other hand, contains terms that dominate in either probe limits
as well an additional mass structure corresponding to the 
first self-force (1SF) corrections. Organizing the results into probe structures and 1SF
structures according to
\begin{equation}
c^{(2,n,i)} = c^{(2,n,i)}_{\text{probe}} + c^{(2,n,i)}_{\text{1SF}}\;,
\end{equation}
truncated to leading order in the dimensional regulator $\epsilon$, yields 
\begin{itemize}
\item Probe structures: 
{\allowdisplaybreaks
\begin{align}
 & c^{(2,0,1)}_{\text{probe}} = -2\pi y\frac{(2y^2-3)(\bar{m}_1^2+\bar{m}_2^2)}{3\epsilon(y^2-1)^2\bar{m}_1\bar{m}_2}\;, \nonumber \\
 & c^{(2,1,1)}_{\text{probe}}= -2\pi\frac{(2y^2-1)(3\bar{m}_1^2+\bar{m}_2^2)}{3\epsilon(y^2-1)^2\bar{m}_1\bar{m}_2}\;,  \\
 & c^{(2,2,1)}_{\text{probe}} = - c^{(2,2,3)}_{\text{probe}} = -\pi y\frac{(10y^2-9)\bar{m}_1^2+(2y^2-1)\bar{m}_2^2}{3\epsilon(y^2-1)^2\bar{m}_1\bar{m}_2}\;, \nonumber \\
 & c^{(2,2,2)}_{\text{probe}} = 2\pi y\frac{(7 y^2-6)\bar{m}_1^2 +y^2\bar{m}_2^2}{3 \epsilon (y^2-1)^3\bar{m}_1 \bar{m}_2}\;, \nonumber 
\end{align}
}

\item 1SF structures:
{\allowdisplaybreaks
\begin{align}
 & c^{(2,0,1)}_{\text{1SF}} = \frac{4\pi(y^4 - 3y^2 + 3)}{3\epsilon(y^2-1)^2}\;, \nonumber \\
 & c^{(2,1,1)}_{\text{1SF}} = \frac{4\pi y(y^2-3)}{3\epsilon(y^2-1)^2}-\frac{4\pi}{\epsilon(y^2-1)^{3/2}}\arcosh(y)\;, \nonumber \\
 & c^{(2,2,1)}_{\text{1SF}} = \frac{2\pi(y^4+1)}{3\epsilon(y^2-1)^2}- 
  \frac{2\pi y(2y^2-1)}{\epsilon(y^2-1)^{5/2}}\arcosh(y)\;, \\
 & c^{(2,2,2)}_{\text{1SF}} = -\frac{\pi(2y^4-11y^2-10)}{3\epsilon(y^2-1)^3} + \frac{\pi y(4y^2-9)}{\epsilon(y^2-1)^{7/2}}\arcosh(y)\;, \nonumber \\
 & c^{(2,2,3)}_{\text{1SF}} =-\frac{\pi(2y^4-3y^2-4)}{3\epsilon(y^2-1)^2} + \frac{\pi y(4y^2-5)}{\epsilon(y^2-1)^{5/2}}\arcosh(y)\; \nonumber.
 \end{align}
}
\end{itemize}
Note that the $1/\epsilon$ pole in the above results are combined with
the factor $(-q^2)^{-2\epsilon}$ in Eq.~\eqref{eq:ampForm}
to generate a Laurent expansion $1/\epsilon - 2 \log(-q^2)$, where the leading divergent term drops out of classical physics
because it involves no non-analytic dependence on $q$.

The one-loop amplitude is well known to exhibit a \textit{spin-shift symmetry}
for gravity~\cite{Aoude:2022trd, Bern:2022kto, Bern:2023ity}, in which $(q
\cdot a)^2$ and $q^2 a^2$ always appear in the following combination 
\begin{equation}
  (q \cdot a)^2 - q^2 a^2\;.
\end{equation}
The consequence is that the amplitude is invariant under a spin shift of the 
form
\begin{equation}
 a^\mu \rightarrow a^\mu +  \xi q^\mu\;,
\end{equation}
for arbitrary $\xi$. For $L=1$ this property is explicitly manifest in our results 
through
\begin{equation}
  c^{(L,2,1)}_{\text{probe}} = -c^{(L,2,3)}_{\text{probe}}\;,
  \label{eq:coeffsHiddenRelation}
\end{equation}
as expected. Furthermore, for the first time, we observe
the above relation holds at two loops for either probe limits,
$\bar m_1 \gg \bar m_2$ or $\bar m_2 \gg \bar m_1$, up to quadratic order in spin. 
The spin-shift symmetry is broken by the 1SF
corrections, which plays a role at two-loops and beyond. We will see
in the following subsection that the symmetry is also present in this same fashion 
in the corresponding gravity amplitude.

\subsection{Gravity}
\label{sec:ScatteringAmplitudesGrav}
The dynamics of a massive scalar and massive vector boson minimally coupled 
to Einstein gravity is given by
\begin{equation}
 \mathcal{L} = \mathcal{L}_\textrm{EH} + \mathcal{L}_\textrm{GF} + 
 \mathcal{L}_\textrm{scalar} + \mathcal{L}_\textrm{vector} + \ldots \;,
\end{equation}
where we ignore higher dimensional operators~\cite{Donoghue:1994dn}. 
The pure gravitational interactions are described by the Einstein-Hilbert (EH) 
Lagrangian, which is given by
\begin{equation}
 \mathcal{L}_\textrm{EH} = -\frac{2}{\kappa^2}\sqrt{-g}~R\;,
\end{equation}
where $g = \det(g_{\mu\nu})$ is the determinant of the metric
$g_{\mu\nu}$, $\kappa^2 = 32\pi G$ is a coupling related to Newton's constant $G$, 
and $R$ is the Ricci scalar. Furthermore, we work in the
weak field approximation by expanding the quantum fluctuations of the metric
tensor around a flat background spacetime metric
\begin{equation}
 g_{\mu\nu} = \eta_{\mu\nu} + \kappa h_{\mu\nu}\;,
\end{equation}
where $h_{\mu\nu}$ is the dynamical graviton quantum field. The quantization
procedure of the graviton field $h_{\mu\nu}$ requires to supplement the
Lagrangian with a gauge-fixing term, where we have chosen the commonly used
\textit{linearized harmonic gauge}. The corresponding Lagrangian term is explicitly
given by
\begin{equation}
 \mathcal{L}_\textrm{GF} = \eta^{\mu\nu} \left(\partial^\lambda h_{\mu\lambda} 
- \frac{1}{2}\partial_\mu h^\lambda_{~\lambda}\right) \left(\partial^\lambda 
 h_{\nu\lambda} - \frac{1}{2}\partial_\nu h^\lambda_{~\lambda}\right)\;,
 \label{eq:gravityLag}
\end{equation}
where indices have been contracted using the flat spacetime metric
$\eta_{\mu\nu}$. In this gauge the $D_s$-dimensional graviton propagator takes the
form~\cite{Bern:2002kj}
\begin{equation}
 P^{\mu\nu,\alpha\beta}(k) = \frac{i}{k^2+i\epsilon}~\frac{1}{2}\left[
 \eta^{\mu\alpha}\eta^{\nu\beta} + \eta^{\mu\beta}\eta^{\nu\alpha} -
 \frac{2}{D_s-2}\eta^{\mu\nu}\eta^{\alpha\beta}\right]\;.
 \label{eq:grav_prop}
\end{equation}
At last, we include massive matter fields for spin $s=0,1$ by 
coupling these minimally to gravity~\cite{Holstein:2008sx}. The spin-$0$ 
massive scalar field $\phi$ is included using
\begin{equation}
 \mathcal{L}_\textrm{scalar} = \sqrt{-g}\left( \frac{1}{2}g^{\mu\nu}
\partial_\mu\phi \partial_\nu\phi - \frac{1}{2}m_\phi^2\phi^2\right)\;, \\
\label{eq:L_scalar}
\end{equation}
while the massive vector field $V_\mu$ is included via
\begin{equation}
 \mathcal{L}_\textrm{vector} = \sqrt{-g}\left( 
 -\frac{1}{4}g^{\mu\alpha}g^{\nu\beta}V_{\alpha\beta}V_{\mu\nu} 
 +\frac{1}{2}m_{\mbox{\tiny V}}^2 g^{\mu\nu} V_\mu V_\nu\right)\;,
\label{eq:L_vector}
\end{equation}
with $V_{\mu\nu}=\partial_\mu V_\nu-\partial_\nu V_\mu$. 
\subsubsection{Loop amplitude calculation}
\label{sec:AnalyticReconGrav}
We reproduce one-loop results by constructing analytic one-loop integrands for scalar-vector scattering,
Eq.~\eqref{eq:phi_phi_V_V}, and scalar-scalar scattering,
Eq.~\eqref{eq:phi_phi_Phi_Phi}, and proceeding as described
in \sectionref{sec:AnalyticUnitarity}.
We bypass the need for gravity Feynman rules using the BCJ double copy
of tree amplitudes~\cite{Bern:2008qj, Bern:2010ue, Bern:2019prr}.
Massless graviton amplitudes in $(D+2)$-dimensions are computed,
followed by dimensional reduction to obtain amplitudes with massive
scalar and vector particles \cite{Bern:2019crd, Johansson:2019dnu,
  Bautista:2019evw}.

As described in \sectionref{sec:NumericalUnitarity} we use the
numerical unitarity method, as implemented in the 
\Caravel{} framework~\cite{Abreu:2020xvt}, to compute all needed
scattering amplitudes, in particular the two-loop ones for which we have no
analytic integrand. We use numerical evaluations of the one-loop amplitudes
to validate the results we obtain from analytic integrands, and
compute the analytic form of the two-loop scalar-vector amplitude by
performing an analytic reconstruction from multiple
numerical evaluations in different kinematic configurations.

The numeric evaluations are performed on a finite number field
with cardinality a 10-digit prime number $p$ (below $2^{31}$). A
parametrization for momentum configurations is built from the
kinematic variables $(\bar m_1, \bar m_2, y, q^2)$ by defining 
the vectors $\bar u_1$, $\bar u_2$, and $q$ according to:
\begin{equation}
  \begin{split}
     \bar u_1 = &\ \left(1, 0, 0, 0\right)\,, \\
     \bar u_2 = &\ \left(\frac{1+x^2}{2x}, 0, 0, \frac{1-x^2}{2x}\right)\,, \\
     q = &\ \left(0, 0, \sqrt{-q^2}, 0\right)\,,
  \end{split}
\end{equation}
and using the relations in \sectionref{sec:KinematicSetUp} to obtain
the momenta $p_i$ ($i=1, 2, 3, 4$). Notice that with this 
parametrization $y=(1+x^2)/(2x)$, and one needs to be able to find 
solutions to the congruence relation $-q^2=n\ (\mathrm{mod}\ p)$. 
We do this by sampling random values for $q^2$ and keeping 
only the configurations for which the Tonelli-Shanks algorithm \cite{MR0371855} returns 
a solution. Effectively, this procedure allows the computation
of a square root in the given finite number field.

\Caravel{} produces two-loop integrands for the amplitude for the 
scalar-vector scattering in Eq.~\eqref{eq:phi_phi_V_V} and for 
scalar-scalar scattering in Eq.~\eqref{eq:phi_phi_Phi_Phi}. Using 
Thiele's interpolation formula, \Caravel{} returns results for fixed
$(\bar m_1, \bar m_2, y)$ values, and with full $\sqrt{-q^2}$ dependence. 
They are written in terms of integrals of the scattering-plane 
tensors introduced in \sectionref{sec:NumericalUnitarity}. Also, the
scalar-vector scattering integrand is given in the tensor-decomposed
form Eq.~\eqref{eq:tensorBasis}.

As in the electrodynamics case, we proceed to perform a soft expansion, IBP 
reduction, and integration. This gives the amplitude as an expansion in the
dimensional regularization parameter $\epsilon = (4-D)/2$, from the
$1/\epsilon^2$ order to the $1/\epsilon$ order, as well as an
expansion over small $|q|$ from order $1/|q|^2$ to $|q|^2$, at
numerical values of $(\bar m_1, \bar m_2, y)$, and the only nontrivial
functions involved, other than rational functions, is
$\operatorname{arcosh}(y)$.

By sampling sufficiently many numerical values for the $(\bar m_1, \bar m_2, y)$
parameters, full analytic expressions can be obtained. Following the
approach of Ref.~\cite{Abreu:2018zmy}, we only reconstruct analytic expressions for
physically meaningful quantities after subtracting of divergences in
the radial-action-like scheme presented in \sectionref{sec:RadialAction}. After
the subtraction, the infrared divergence $\sim 1/\epsilon^2$ and
superclassical divergence $\sim 1/|q|^2, 1/|q|$ disappear. We are left
with an amplitude whose $\epsilon$ expansion starts at
$\mathcal O(1/\epsilon)$, which combined with the Taylor expansion of the
overall factor $(-q^2)^{-2\epsilon}$ generates a non-analytic
$\log(-q^2)$ term responsible for classical interactions. The $|q|$
expansion now includes the classical $\mathcal O(|q|^0)$ and the
(naively) quantum order $\mathcal O(|q|^2)$. While the
$\mathcal O(|q|^0)$ analytic result is fully reconstructed, the only
relevant information in the $\mathcal O(|q|^2)$ terms is the gradient
of the spin-diagonal part with respect to $S^2 = -s(s+1)$, which gives
the spin Casimir terms, as discussed in
\sectionref{sec:SpinInterpolation}. Therefore we only reconstruct analytic
expressions for this gradient. Since the finite remainder of the amplitude has
3 mass structures, we sample 3 different values of $(\bar m_1, \bar
m_2)$ for each value of $y$, with a total of about 30 values of $y$
used to reconstruct the final analytic expressions.
\subsubsection{Classical amplitude and spin-shift symmetry}
\label{sec:AmplitudeGrav}

In the same notation of \sectionref{sec:AmplitudeQED} for electrodynamics, the four-point tree-level, one-loop, and two-loop gravity 
amplitudes for scalar-vector scattering, in the radial-action-like scheme for subtracting superclassical divergences, take the form
\begin{equation}
	\begin{split}
		& \mathcal{M}^{(0)}_{\text{finite}} = \frac{G}{(-q^2)}\sum_{n = 1}^{4} M_n^{(0)}T_n + \mathcal{O}(q^0)\;, \\
		& \mathcal{M}^{(L)}_{\text{finite}} = \frac{G^{1+L}}{(-q^2)^{1-L/2+L\epsilon}}  \sum_{n = 1}^{4} M_n^{(L)} T_n + \text{subleading terms in small } |\boldsymbol{q}|\;.
	\end{split}
\end{equation}
Here, $M_1^{(L)}$ is again separated into two pieces 
\begin{equation}
	M_1^{(L)} = -M_{\text{spin}-0}^{(L)} - q^2M_{\text{gradient}}^{(L)} + \ldots \;,
\end{equation}
and $\ldots$ \ denote truly quantum-suppressed non-gradient terms.
The explicit coefficients can be found in the supplemental material file {\tt finiteAmplitudes.m}.

As before, the finite remainder of the amplitude, written in a form for
massive particles of any spin and truncated at the quadratic order in
spin, is
\begin{equation}
  \mathcal{M}_{\text{finite}}^{(L)} = \frac{G^{1+L}}{(-q^2)^{1-L/2+L\epsilon}} \sum_{n,i}^{} c^{(L,n,i)} Q^{(n,i)}\;.
 \label{eq:GR_finite}
\end{equation}
To avoid cluttering the notation, here we use the same coefficient names
as used before for the electrodynamics results. The context should leave no confusion. 
The tree-level coefficients are
\begin{equation}
	\begin{split}
		& c^{(0,0,1)} = 16\pi(2y^2-1)\bar{m}_1^2\bar{m}_2^2\;, \qquad c^{(0,1,1)} = 32\pi y \bar{m}_1^2\bar{m}_2^2\;, \\
		& c^{(0,2,1)} = 8\pi(2y^2-1)\bar{m}_1^2\bar{m}_2^2\;, \qquad ~~ \hspace{-0.5mm}c^{(0,2,2)} = c^{(0,2,3)} = 0\;.
	\end{split}
\end{equation}
The one-loop coefficients read
\begin{equation}
	\begin{split}
		& c^{(1,0,1)} = 6\pi^2(5y^2-1)(\bar{m}_1+\bar{m}_2)\bar{m}_1^2\bar{m}_2^2\;, \\
		& c^{(1,1,1)} = 2\pi^2 y\frac{(5y^2-3)(4\bar{m}_1 + 3\bar{m}_2)\bar{m}_1^2 \bar{m}_2^2}{(y^2-1)}\;, \\
		& c^{(1,2,1)} = - c^{(1,2,3)} = \pi^2\frac{((95y^4 - 102y^2+15)\bar{m}_1+4(15y^4 - 15y^2 + 2)\bar{m}_2)\bar{m}_1^2\bar{m}_2^2}{4(y^2-1)}\;, \\
		& c^{(1,2,2)} = -\pi^2\frac{((65y^4 - 66y^2+9)\bar{m}_1+2(15y^4 - 12y^2 + 1)\bar{m}_2)\bar{m}_1^2\bar{m}_2^2}{2(y^2-1)^2}\;, 
	\end{split}
\end{equation}
Finally, the two-loop coefficients are organized again into probe structures, which 
dominate in either the $\bar m_1 \gg \bar m_2$ or $\bar m_2 \gg \bar m_1$ limits, 
and 1SF structures, according to
\begin{equation}
c^{(2,n,i)} = c^{(2,n,i)}_{\text{probe}} + c^{(2,n,i)}_{\text{1SF,rational}} + c^{(2,n,i)}_{\text{1SF},\arcosh(y)}\;.
\end{equation}
Here we split rational and $\arcosh$ terms of the 1SF contributions. 
The individual terms are given by
\begin{itemize}
	\item Probe structures: 
		{\allowdisplaybreaks
		\begin{align}
			& c^{(2,0,1)}_{\text{probe}} = \frac{2\pi(64y^6 - 120y^4 + 60y^2 -5)(\bar{m}_1^2+\bar{m}_2^2)\bar{m}_1^2\bar{m}_2^2}{3  \epsilon(y^2-1)^2}\;, \nonumber \\
			& c^{(2,1,1)}_{\text{probe}}= 4\pi y\frac{(16y^4-20y^2+5)(3\bar{m}_1^2+2\bar{m}_2^2)\bar{m}_1^2\bar{m}_2^2}{3 \epsilon(y^2-1)^2}\;, \\
			& c^{(2,2,1)}_{\text{probe}} = - c^{(2,2,3)}_{\text{probe}} = \pi\frac{((128y^6-216y^4+96y^2-7)\bar{m}_1^2+(64y^6-104y^4+44y^2-3)\bar{m}_2^2)\bar{m}_1^2\bar{m}_2^2}{3\epsilon(y^2-1)^2}\;, \nonumber \\
			& c^{(2,2,2)}_{\text{probe}} = -2\pi\frac{((80y^6-132y^4+57y^2-4)\bar{m}_1^2+(32y^6-48y^4+18y^2-1)\bar{m}_2^2)\bar{m}_1^2\bar{m}_2^2}{3 \epsilon(y^2-1)^3}\;, \nonumber 
		\end{align}
		}
	\item 1SF rational terms:
	{\allowdisplaybreaks
	\begin{align}
		& c^{(2,0,1)}_{\text{1SF,rational}} =  4\pi y\frac{(36y^6-114y^4+132y^2-55)\bar{m}_1^3 \bar{m}_2^3}{3 \epsilon(y^2-1)^2}\;, \nonumber \\
		& c^{(2,1,1)}_{\text{1SF,rational}} = 4\pi \frac{(36y^6 - 156y^4 + 84y^2 +41)\bar{m}_1^3 \bar{m}_2^3}{3\epsilon(y^2-1)^2}\;, \nonumber \\
		& c^{(2,2,1)}_{\text{1SF,rational}}  = 2\pi y\frac{(180y^6-1242y^4+104y^2+213)\bar{m}_1^3 \bar{m}_2^3}{15 \epsilon(y^2-1)^2}\;, \\
		& c^{(2,2,2)}_{\text{1SF,rational}}  = -2\pi y\frac{(138y^6-1769y^4+1394y^2+2122)\bar{m}_1^3 \bar{m}_2^3}{15 \epsilon(y^2-1)^3}\;,\nonumber  \\
		& c^{(2,2,3)}_{\text{1SF,rational}}  = -2\pi y\frac{(186y^6-1257y^4+446y^2+1005)\bar{m}_1^3 \bar{m}_2^3}{15 \epsilon(y^2-1)^2}\;. \nonumber
	\end{align}
	}
	\item 1SF $\arcosh(y)$ terms:
	\begin{equation}
		\begin{split}
			& c^{(2,0,1)}_{\text{1SF},\arcosh(y)} =  8\pi \frac{(4y^4-12y^2-3)\bar{m}_1^3 \bar{m}_2^3}{ \epsilon\sqrt{y^2-1}}  \arcosh(y)\;, \\
			&  c^{(2,1,1)}_{\text{1SF},\arcosh(y)} = 16\pi y\frac{(2y^4-11y^2-6)\bar{m}_1^3\bar{m}_2^3}{ \epsilon(y^2-1)^{3/2}} \arcosh(y)\;, \\
			& c^{(2,2,1)}_{\text{1SF},\arcosh(y)} = 4\pi y^2\frac{(4y^6 -36y^4+y^2+6)\bar{m}_1^3\bar{m}_2^3}{\epsilon(y^2-1)^{5/2}} \arcosh(y)\;, \\
			& c^{(2,2,2)}_{\text{1SF},\arcosh(y)} = -2\pi\frac{(8y^8 -104y^6+35y^4+162y^2+24)\bar{m}_1^3\bar{m}_2^3}{ \epsilon(y^2-1)^{7/2}} \arcosh(y)\;,  \\
			& c^{(2,2,3)}_{\text{1SF},\arcosh(y)}= -2\pi\frac{(8y^8 -72y^6+11y^4+66y^2+12)\bar{m}_1^3\bar{m}_2^3}{ \epsilon(y^2-1)^{5/2}} \arcosh(y)\;.
		\end{split}
	\end{equation}
\end{itemize}
As explained for the amplitudes in electrodynamics, the $1/\epsilon$ pole
is responsible for the finite $\log(-q^2)$ terms that
ultimately contribute to classical physics. The spin-shift symmetry
is, once again, manifest for all probe structures, as the linear
relation Eq.~\eqref{eq:coeffsHiddenRelation} is satisfied, but is
broken by the 1SF correction.


\section{Aligned-spin observables and radial action}
\label{sec:Observables}
We use the radial-action-like subtraction scheme to compute
aligned-spin observables for both electrodynamics and gravity, focusing on the
scattering angle.
The aligned-spin limit fully captures the spin Casimir terms and
tests our method of spin interpolation for extracting such terms from
fixed-spin scattering amplitudes. The non-Casimir terms at $\mathcal{O}(G^3S^2)$ have
already been computed from scattering amplitudes in the earlier work \cite{FebresCordero:2022jts}.

To this end, the radial action is determined from the finite remainder of the amplitude through the
following Fourier transformation
\begin{equation}
 I_r^{(L)} = \int \hat{d}^4 q\, \hat{\delta}(\bar{p}_1 \cdot q) \hat{\delta}(-\bar{p}_2 \cdot q) 
 e^{i b \cdot q} \mathcal{M}^{(L)}_{\text{finite}}\;,
 \label{eq:RadialActionFourierTransform}
\end{equation}
where $\mathcal{M}^{(L)}_{\text{finite}}$ is the finite remainder of 
the amplitude in terms of spin, which is given in \equationref{eq:QED_finite} for QED and in 
\equationref{eq:GR_finite} for gravity. Furthermore, we absorb factors of $2
\pi$ in both integrand measures and delta functions using the notation
\begin{equation}
 \hat{d}^{D} x \equiv \frac{d^{D} x}{(2\pi)^D}\;, \qquad \hat{\delta}(x) \equiv 2\pi \delta(x)\;.
\end{equation}
As a consequence of the spin structures, we need to compute tensor integrals
of the general form
\begin{equation}
    f^{\mu_1 \cdots \mu_n}_\beta = \int \hat{d}^4 q\, \hat{\delta}(\bar{p}_1 \cdot q) \hat{\delta}(-\bar{p}_2 \cdot q) 
 	e^{i b \cdot q} q^{\mu_1}\ \cdots q^{\mu_n} (-q^2)^{-\beta}\;,
 	\label{eq:GeneralRankN_noI}
\end{equation} 
where the necessary integrals, alongside a discussion of how to evaluate them,
can be found in \appendixref{appendix:EvaluatingRankedIntegrals}. After the
relevant integrals are performed, we specialize to the aligned-spin case by
imposing
\begin{equation}
   a \cdot \bar{u}_i = a \cdot \hat{b} = 0\;, \quad 
   \epsilon^{\mu \nu \rho \sigma} \bar{u}_{1\nu} \bar{u}_{2\rho} a_\sigma =  |a| \sqrt{y^2-1} \ \hat{b}^\mu\;, \quad 
   a \cdot a = -|a|^2\;,
\end{equation}
where $\bar{u}_i$ span the equatorial plane (see Eq.~\eqref{eq:velvectorsproperties}), we align the normalized spin vector, 
$a$, with the angular momentum, and we define the normalized impact parameter 
$\hat{b}^\mu = b^\mu/|b|$. The aligned-spin scattering angle can then be computed
using Eq.~\eqref{eq:angleradaction},
with $J = |\boldsymbol{p}| |\boldsymbol{b}|$ the angular momentum in the asymptotic past, 
together with
\begin{equation}
 |b|^2 = -b^2 = \boldsymbol{b}^2\;, \qquad 
 |\boldsymbol{p}| = \frac{\bar{m}_1 \bar{m}_2 \sqrt{y^2 - 1}}{\sqrt{\bar{m}_1^2 + \bar{m}_2^2 + 2\bar{m}_1 \bar{m}_2y}} 
 = \frac{\bar{m}_1 \bar{m}_2 \sqrt{y^2 - 1}}{\sqrt{s}}\;,
\end{equation} 
where $s = (p_1 + p_2)^2$ is the usual Mandelstam variable. The scattering angle in  Eq. \eqref{eq:angleradaction} is expanded in the coupling constant
\begin{equation}
  \chi = g\,\chi^{(0)} + g^2\chi^{(1)} + g^3\chi^{(2)} + \mathcal{O}(g^4)\;, 
  \label{eq:ScatteringAngle_Expansion}
\end{equation}
where $g$ denotes the coupling constant, i.e. $\alpha_{\text{eff}}$ for QED 
and $G$ for gravity.

In practice we need to perform derivatives of the radial action with 
respect to $J$ to acquire the scattering angle. To do this, we first realise that the action of the 
differentiation in Eq.~\eqref{eq:angleradaction} can be recast as
\begin{equation}
	\frac{\partial}{\partial J} = \frac{1}{|\boldsymbol{p}|} \frac{\partial}{\partial |\boldsymbol{b}|} \iff \frac{1}{|\boldsymbol{p}|} \frac{b_\mu}{|\boldsymbol{b}|} \frac{\partial}{\partial b_\mu},
\end{equation}
such that we first differentiate Eq.~\eqref{eq:RadialActionFourierTransform} and then evaluate 
the tensor integrals of the form Eq.~\eqref{eq:GeneralRankN_noI}. To be more concrete, let us compute 
the electrodynamics scattering angle at the leading order. From the tree-level amplitude in scalar QED
{\allowdisplaybreaks
\begin{align}
	\chi^{(0)}_{\text{SQED}} & = - \frac{\partial}{\partial J} I_{r,\text{SQED}}^{(0)} \nonumber \\
	& = \frac{16y\pi \bar{m}_1 \bar{m}_2}{|\boldsymbol{p}|} \frac{b_\mu}{|\boldsymbol{b}|} \frac{\partial}{\partial b_\mu}  \int \hat{d}^4 q\, \hat{\delta}(\bar{p}_1 \cdot q) \hat{\delta}(-\bar{p}_2 \cdot q) 
 	e^{i b \cdot q} \frac{1}{(-q^2)} \nonumber \\
 	& = \frac{16y\pi \bar{m}_1 \bar{m}_2}{|\boldsymbol{p}|} \frac{b_\mu}{|\boldsymbol{b}|}\int \hat{d}^4 q\, \hat{\delta}(\bar{p}_1 \cdot q) \hat{\delta}(-\bar{p}_2 \cdot q) 
 	e^{i b \cdot q} \frac{i q^\mu}{(-q^2)} \\
 	& = -\frac{2y}{|\boldsymbol{p}||\boldsymbol{b}|\sqrt{y^2-1}} \nonumber\\
 	& = -\frac{2y}{J\sqrt{y^2-1}} \nonumber \;,
\end{align}
}
which is the well-known leading-order scattering angle for electrodynamics once multiplied by 
$\alpha_{\textrm{eff}}= q_\phi q_{\mbox{\tiny V}} \alpha$.
\subsection{Observables in electrodynamics}
\label{sec:ObservablesQED}
As we have determined the finite classical limit, we may drop the barred
notation $\bar{m}_i$ for finite classical observables, since the difference
between barred and unbarred variables is quantum suppressed. Here we state the scattering angle
in electrodynamics up to two loops, i.e.\ $\mathcal O(\alpha_{\text{eff}}^3)$, and quadratic order in spin. At leading order we have
\begin{equation}
 \chi^{(0)}_{\text{aligned-spin}}  = - \frac{2y}{J\sqrt{y^2-1}} + \frac{2 m_1m_2 \sqrt{y^2-1}}{J^2 \sqrt{s}}|a|   - \frac{2m_1^2 m_2^2 y \sqrt{y^2-1}}{J^3 s} |a|^2\;,
\end{equation}
at next-to-leading order we find
\begin{equation}
\begin{split}
    \chi^{(1)}_{\text{aligned-spin}}   = \frac{\pi (m_1 + m_2)}{2 J^2 \sqrt{s}} & - \frac{\pi y (2m_1+m_2)m_1m_2}{J^3 s} |a|  \\
    & + \frac{3\pi ((5y^2-3)m_1 + 2y^2 m_2)m_1^2m_2^2}{4 J^4 s^{3/2}} |a|^2\;,
\end{split}
\end{equation}
and at next-to-next-to-leading order we have 
\begin{equation}
\begin{split}
    \chi^{(2)}_{\text{aligned-spin}} = & -\frac{2y(2y^2-3)(m_1^2+m_2^2)}{3J^3s(y^2-1)^{3/2}} 
    + \frac{4(y^4-3y^2+3)m_1m_2}{3J^3  s (y^2-1)^{3/2}}\\
    & +\frac{2(2y^2-1)(3m_1^2 + m_2^2)m_1m_2}{J^4 s^{3/2} \sqrt{y^2-1}} |a|
    - \frac{4y(y^2-3)m_1^2m_2^2}{J^4 s^{3/2} \sqrt{y^2-1}}|a| \\
    & + \frac{12 m_1^2m_2^2}{J^4 s^{3/2}}  \arcosh(y) |a|
  	 -\frac{4y((10y^2-9)m_1^4 m_2^2 + (2y^2-1)m_1^2 m_2^4)}{J^5 s^2 \sqrt{y^2-1}}|a|^2 \\
  	& + \frac{8(y^4-y^2-1)m_1^3m_2^3}{J^5 s^2 \sqrt{y^2-1}}|a|^2 
      - \frac{48 m_1^3m_2^3 y }{J^5 s^2}\arcosh(y)|a|^2\;,
\end{split}
\end{equation}
where we identify the angular momentum at past infinity as 
\begin{equation}
 J = |\boldsymbol{b}\times\boldsymbol{p}| = |\boldsymbol{b}||\boldsymbol{p}|\;.
\end{equation}
To validate our results, we further specialize our analysis to the 
limiting case of a probe scalar ($m_1 \gg m_2$)
\begin{equation}
\begin{split}
    & \chi^{(0)}_{\text{aligned-spin,probe}}  = - \frac{2y}{J\sqrt{y^2-1}} + \frac{2m \sqrt{y^2-1}}{J^2} |a|   - \frac{2 m^2 y \sqrt{y^2-1}}{J^3}|a|^2 \;, \\ 
    & \chi^{(1)}_{\text{aligned-spin,probe}}   = \frac{\pi}{2 J^2} - \frac{2\pi y m }{J^3}|a|  + \frac{3 \pi (5y^2-3) m^2}{4 J^4}|a|^2 \;, \\ 
    & \chi^{(2)}_{\text{aligned-spin,probe}}   = -\frac{2y(2y^2-3)}{3J^3 (y^2-1)^{3/2}} +\frac{6 (2y^2-1)m}{J^4\sqrt{y^2-1}}|a|  -\frac{4 y (10y^2-9) m^2 }{J^5\sqrt{y^2-1}}|a|^2\;,
\end{split}
\label{eq: QEDScatteringAmmplitudeAngle}
\end{equation}
where we relabel $m_2 = m$.

\subsubsection*{Validation: Classical aligned-spin probe limit}
\label{sec:CrossCheckQED}
Observables for spinning two-body interactions in electrodynamics
have been computed using scattering amplitudes 
to $\mathcal{O}(\alpha S^2)$ in Ref.~\cite{Maybee:2019jus}, and to 
$\mathcal{O}(\alpha^2 S^2)$ in Ref.~\cite{Menezes:2022tcs} using the
massive spinor-helicity formalism.
As there are no results in the literature for such
observables beyond
one loop, the simplest cross-check we can make is to solve the classical
dynamics of a charged probe scalar mass in a root-Kerr 
background \cite{Monteiro:2014cda, Arkani-Hamed:2019ymq} which is the
electromagnetic field in the presence of a spinning disc of charge. The root-Kerr
four-vector potential solution relevant for the dynamics of a probe mass confined to the equatorial
plane, i.e. the $(x,y)$-plane, reads
\begin{equation}
    A_\mu \big \rvert_{z = 0} = \frac{\alpha}{e \tilde{r}} \left(1, \frac{x\tilde{r} + ay}{r^2}, \frac{y\tilde{r} - ax}{r^2}, 0\right)\;, 
    \label{eq: rootKerr}
\end{equation}
where $\tilde{r} = \sqrt{r^2 - a^2}$ and $a = |a|$ is the spin associated to the
rotating ring of charge. The dynamics for the worldline of an electrically
charged point particle of mass $m$ reads 
\begin{equation}
 S = \int d\tau \left(\frac{m}{2} \dot{X}_\mu \dot{X}^\mu + \sqrt{4\pi \alpha} A_\mu \dot{X}^\mu \right)\;, \quad 
 X^\mu = (t,x,y,z)\;, \quad 
 \dot{X}^\mu = \frac{dX^\mu}{d\tau}\;,
\end{equation}
which we find by gauge fixing the usual polynomial action. Then, one can determine
the Noether currents corresponding to time translation and rotation invariance,
which results in equations for energy and angular momentum conservation
\begin{equation}
  m \dot{X}^0 + eA^0 = \text{constant}\;, \qquad (m \dot{X}_i + e A_i) \Omega_{ij} X_j = \text{constant}\;,
\end{equation}
where
\begin{equation}
  \Omega_{ij} = \begin{pmatrix}
    0 & -1 \\
    1 & \phantom{-}0
  \end{pmatrix}\;,
\end{equation}
is the infinitesimal rotation matrix. Taking
the initial conditions to be that of the probe mass at past infinity with
velocity $v_{\infty}$, energy and angular momentum conservation allow us to
determine the total velocity, $v$, and angular velocity, $v_{\theta}$, as a
function of radial distance, respectively. More specifically, this is done by 
solving
\begin{equation}
  \sqrt{v^2 + 1} - \sqrt{v_{\infty}^2 + 1} + \frac{\alpha}{m \tilde{r}} = 0\;, \qquad 
  m (\boldsymbol{r} \times \boldsymbol{v})_z + e (\boldsymbol{r} \times \boldsymbol{A})_z = m b v_\infty\;,
\end{equation}
where the impact parameter $b = |\boldsymbol{b}|$ is related to the angular momentum at infinity 
via $J = m v_{\infty} b$. To this end, we find
\begin{equation}
	\begin{split}
		& \begin{split}
			v(r,a) = v_{\infty} - \frac{(2r^2+a^2)\sqrt{1+v_{\infty}^2}}{2mr^3 v_{\infty}}\alpha 
			& - \frac{(r^2+a^2)}{2m^2 r^4 v_{\infty}^3}\alpha^2 \\
			& - \frac{(2r^2+3a^2)\sqrt{1+v_{\infty}^2}}{4m^3 r^5 v_{\infty}^5} \alpha^3 + \mathcal{O}(a^3,\alpha^4)\;, 
			\end{split} \\
		& v_{\theta}(r,a) = \frac{b v_\infty}{r} - \frac{a }{m r^2}\alpha + \mathcal{O}(a^3)\;,
	\end{split}
\end{equation}
where the radial velocity, $v_{r}$, is trivially fixed through 
$v^2 = v_{r}^2 + v_{\theta}^2$. Then, we can determine the scattering angle as
\begin{equation}
  \chi = 2\theta_{r_{\text{min}} \rightarrow \infty} - \pi\;,
\end{equation} 
together with
\begin{equation}
    \theta_{r_{\text{min}} \rightarrow \infty} = \int_{r_{\text{min}}}^{\infty} dr \frac{d\theta}{dr} = \int_{r_{\text{min}}}^{\infty} dr \frac{v_\theta}{r v_r}\;,
\end{equation}
where $r_{\text{min}}$ follows by solving $v_r = 0$. Finally, integrating and 
organizing the scattering angle as in Eq. \eqref{eq:ScatteringAngle_Expansion}, we 
find
\begin{equation}
\begin{split}
    & \chi_{cl}^{(0)}  = - \frac{2y}{J\sqrt{y^2-1}} + \frac{2m \sqrt{y^2-1}}{J^2}a  - \frac{2 m^2 y \sqrt{y^2-1}}{J^3} a^2\;, \\ 
    & \chi_{cl}^{(1)}  = \frac{\pi}{2 J^2} - \frac{2\pi ym }{J^3}a + \frac{3\pi (5y^2-3) m^2 }{4 J^4}a^2\;, \\ 
    & \chi_{cl}^{(2)}  = -\frac{2y(2y^2-3)}{3J^3 (y^2-1)^{3/2}} + \frac{6 (2y^2-1)m}{J^4\sqrt{y^2-1}} a-\frac{4y (10y^2-9) m^2 }{J^5\sqrt{y^2-1}}a^2\;,
\end{split}
\end{equation}
where we identify
\begin{equation}
    v_\infty = \sqrt{y^2 - 1}\;, \qquad b = \frac{J}{m \sqrt{y^2-1}}\;.
\end{equation}
This is in exact agreement with \equationref{eq: QEDScatteringAmmplitudeAngle}.
\subsection{Observables in gravity}
\label{sec:ObservablesGrav}

Here we state the scattering angle in gravity up to two loops, i.e.\
$\mathcal O(G^3)$, and quadratic order in spin. 
At leading order we find
\begin{equation}
	\begin{split}
		\chi^{(0)}_{\text{aligned-spin}} = \frac{2(2y^2-1)m_1 m_2}{J\sqrt{y^2-1}} & - \frac{4y m_1^2 m_2^2 \sqrt{y^2-1}}{J^2 \sqrt{s}}|a| \\
		& + \frac{2m_1^3m_2^3 \sqrt{y^2-1}(2y^2-1)}{J^3 s} |a|^2\;,
	\end{split}
\end{equation}
at next-to-leading order we find
\begin{equation}
\begin{split}
    \chi^{(1)}_{\text{aligned-spin}} = & \ \frac{3\pi (5y^2-1)(m_1 + m_2)m_1^2 m_2^2}{4J^2 \sqrt{s}} - \frac{\pi y (5y^2-3)(4m_1+3m_2)m_1^3m_2^3}{2J^3 s} |a|  \\
    	& +\frac{3\pi ((95y^4-102y^2+15)m_1 + (60y^4-60y^2 + 8) m_2)m_1^4m_2^4}{16 J^4 s^{3/2}} |a|^2\;,
\end{split}
\end{equation}
in agreement with well known results in the literature,
e.g.~\cite{Kosmopoulos:2021zoq, Liu:2021zxr, Bern:2022kto,
  Aoude:2022trd, Aoude:2022thd}.
Finally, at next-to-next-to-leading order we find
{\allowdisplaybreaks
\begin{align}
    & \chi^{(2)}_{\text{aligned-spin}} = \frac{2(64y^6-120y^4 + 60y^2 -5)(m_1^2+m_2^2)m_1^3 m_2^3}{3J^3 s (y^2-1)^{3/2}} \nonumber \\
    & + \frac{4y(36y^6-114y^4+132y^2-55)m_1^4m_2^4}{3 J^3 s(y^2-1)^{3/2}} - \frac{8(4y^4-12y^2-3)m_1^4m_2^4}{J^3 s}\arcosh(y) \nonumber \\
    & - \frac{4y(16y^4-20y^2+5)(3m_1^2 + 2 m_2^2)m_1^4 m_2^4}{J^4  s^{3/2}\sqrt{y^2-1}}|a| \nonumber \\
    & - \frac{4y(36y^6-156y^4+84y^2+41)m_1^5m_2^5}{J^4 s^{3/2}\sqrt{y^2-1}}|a| \\
    & + \frac{48y(2y^4 - 11y^2 - 6)m_1^5m_2^5}{J^4 s^{3/2}}\arcosh(y)|a| \nonumber \\
    & + \frac{4((128y^6 - 216y^4 + 96 y^2 - 7)m_1^2 + (64y^6 - 104y^4 + 44y^2 -3)m_2^2)m_1^5 m_2^5}{J^5 s^2 \sqrt{y^2-1}}|a|^2 \nonumber \\
    & + \frac{8 y(184 y^6 - 1252 y^4 + 332 y^2 + 741)m_1^6 m_2^6}{5 J^5 s^2 \sqrt{y^2-1}}|a|^2 \nonumber - \frac{192(y^6 - 8 y^4 -7y^2 -1)m_1^6 m_2^6}{J^5 s^2}\arcosh(y)|a|^2\; \nonumber .
\end{align}
}

\noindent
This is in agreement with Ref.~\cite{Jakobsen:2022fcj} after converting their
result to the scattering angle in the center-of-mass frame.


\section{Conclusion}
\label{sec:Conclusion}
In this paper we have developed a method to resolve the ambiguity in
computing spin Casimir terms for the dynamics of binary systems
involving spinning black holes from scattering amplitudes of
fixed-spin theories.
The ambiguity originates from the fact that such terms are just
numbers, $S^2 = -s(s+1)$ for spin-$s$ particles. When combined with
the accompanying factor $q^2$, i.e.\ the squared momentum transfer of
order $\hbar^2$, these terms are naively indistinguishable from the
quantum-suppressed spin-independent terms of the amplitude.  We have
presented a new method, \emph{spin interpolation}, to cleanly extract
the Casimir terms by computing the gradient of the spin-diagonal part
of the amplitude with respect to $S^2$.  We have demonstrated the
validity of the method by calculating the two-body dynamics of a
spinless object and a spinning object, up to the quadratic order in
spin, using spin-$0$ and spin-$1$ amplitudes in both electrodynamics 
and gravity. This has been done up to the two-loop level. This completes the 
earlier work \cite{FebresCordero:2022jts} involving some of the 
authors here, and together constitute the first calculation of spin effects 
in gravitational binary dynamics at the third post-Minkowskian order
using scattering amplitudes, previously only calculated by worldline
approaches \cite{Jakobsen:2022fcj, Jakobsen:2022zsx}.  The method is
completely general, and there is no obstruction in future applications
to higher orders in spin, requiring a larger tower of amplitudes with
different spin representations in order to extract the full classical
information.

The method relies on spin universality -- the amplitude is written in
terms of abstract spin operators that can act on an arbitrary massive
spin representation. In fact, the correctness of our results
demonstrates a very strong form of spin universality which applies to
both classical and quantum-suppressed terms of the amplitude and
applies to finite spin representations without explicitly taking a
large-spin limit. In the future, it would be interesting to explore
the spin interpolation procedure at an earlier stage in the
calculation and write the four- and five-point Compton amplitudes
(used in generalized unitarity cuts) in terms of abstract spin
operators, and compare with the arbitrary-spin Lagrangian formalism of
Refs.~\cite{Bern:2020buy, Kosmopoulos:2021zoq, Bern:2022kto,
  Bern:2023ity}.

We have extended the radial-action-like subtraction scheme
\cite{Bern:2021dqo, Bern:2021yeh, Bern:2024adl} -- in which one
effectively deletes divergent master integrals -- to compute
aligned-spin observables. The amplitude-action relation, established
in the spinless case, states that the finite remainder of the
amplitude in this subtraction scheme coincides with the radial
action. We have observed that the amplitude-action relation remains
valid in the presence of one spinning body in the aligned-spin limit,
up to the third order in the electromagnetic/gravitational coupling
constant. In fact, this is the subtraction scheme we used to calculate
the spin Casimir terms in the scattering angles, finding agreement with
classical equations of motions in a root-Kerr background in the case
of electrodynamics and agreement with known results from worldline
calculations in the case of gravity. We leave it to future work to
develop this scheme for misaligned spin observables.

Surprisingly, when specializing to the limit where either the spinless
body or the spinning body becomes a probe object, the finite remainders
of the two-loop amplitude in both gravity and electrodynamics exhibit a
spin-shift symmetry, which was thought to be an accidental property at
one loop \cite{Aoude:2022trd, Bern:2022kto, Bern:2023ity}. This is the
first instance of this property being present beyond one-loop, and it
is natural to conjecture that the property, while tied to low orders
in spin \cite{Bautista:2022wjf} and the probe limit, otherwise holds
at all orders in the coupling constant. Note that the entire one-loop
amplitude is a sum of the two probe limits, while the first-self-force
correction beyond the probe limit, starting at two loops, violates the
spin-shift symmetry. In light of the domain of validity of the
symmetry, we further conjecture that the symmetry is a consequence of
the integrability of probe black hole motions in a Kerr background in
the case of a non-spinning probe \cite{Carter:1968rr} or a spinning
probe at the first few orders in the probe spin
\cite{rudiger1981conserved, rudiger1983conserved, Compere:2023alp}, as
well as the electrodynamics analog of integrable motion in a root-Kerr
background \cite{Ball:2023xnr}. Ref.~\cite{Gonzo:2024zxo} proposed a
simple factorized form for the radial action for a probe object in a
Kerr background as a consequence of integrability, and this can be
taken as a hint that the relevant scattering amplitudes also exhibit
hidden symmetry.  Further evidence for the connection with
integrability comes from the fact that when one includes generic
spin-induced multipole moments beyond the Kerr black hole case
(e.g. for neutron stars), Refs.~\cite{Aoude:2022trd, Bern:2022kto}
observe violation of the spin-shift symmetry while
Ref.~\cite{Compere:2023alp} observes violation of the integrability of
Kerr orbits, again showing perfect alignment between the two sides. We
leave it to future work to directly establish this connection.

\section*{Acknowledgments}
We thank Rafael Aoude, Fabian Bautista, Lucile Cangemi, Riccardo Gonzo, Henrik Johansson,
Jung-Wook Kim, Dimitrios Kosmopoulos and Irene Roman for insightful discussions.
We especially thank Guanda Lin for collaboration during the initial
phases of the project and for insightful discussions.
Some of the collaboration and discussions took place at the Gravitational Waves meet 
Amplitudes in the Southern Hemisphere Program at the ICTP-SAIFR
and the QCD Meets Gravity 2023 conference at CERN.
We extensively use the Mathematica package {\tt FeynCalc}
\cite{Shtabovenko:2023idz} in some of the calculations.
D.A.\ is supported by a STFC studentship.
The work of F.F.C. is supported in part by the U.S. Department of Energy
under grant DE-SC0010102.
M.K.\ is supported by the DGAPA-PAPIIT grant IA102224 (``Iluminando agujeros
negros") and the PIIF at UNAM.
This work has been made possible in part through the support of the FSU
Council on Research and Creativity (``Black Holes Under the Microscope";
SEED Grant, 2023).
M.S.R.\ is supported by the U.S. Department of Energy (DOE)
under award number DE-SC0009937.
M.Z.’s work is supported in part by the U.K.\ Royal Society through
Grant URF\textbackslash R1\textbackslash 20109.
For the purpose of open access, the authors
have applied a Creative Commons Attribution (CC BY) license to any
Author Accepted Manuscript version arising from this submission.


\appendix
\section{Conventions}
\label{appendix:Conventions}
We adopt the mostly negative metric signature $\eta = \operatorname{diag}(1,-1,\ldots, -1)$
together with $\epsilon_{0123}=+1$. In usual fashion, total symmetrization and anti-symmetrization is defined as 
\begin{equation}
	\begin{split}
		& X^{(\mu_1} \cdots X^{\mu_n)} = \frac{1}{n!} ( X^{\mu_1} X^{\mu_2}\cdots X^{\mu_n} + X^{\mu_2} X^{\mu_1} \cdots X^{\mu_n}+ \cdots)\;, \\
		& X^{[\mu_1} \cdots X^{\mu_n]} = \frac{1}{n!} ( X^{\mu_1} X^{\mu_2}\cdots X^{\mu_n} - X^{\mu_2} X^{\mu_1} \cdots X^{\mu_n}+ \cdots)\;.
	\end{split}
\end{equation}
Finally, we define scattering amplitudes with the normalization that
they are equal to $(-i)$ times the sum of Feynman diagrams.
Amplitudes in QED and gravity are denoted 
by $\mathcal{A}$ and $\mathcal{M}$, respectively, and we use the latter 
for general amplitude discussions.


\section{Feynman rules for scalar-vector QED}
\label{appendix:ScalarVectorQED}
Here we outline the Feynman rules that follow from \equationref{eq:
EMLagrangian}. The propagators read 
\begin{equation}
\begin{aligned}
    \raisebox{-0.25cm}{\includegraphics[scale=1.1]{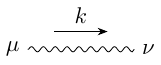}}
    \ & = \ \frac{i}{k^2 + i\epsilon}\left(-\eta^{\mu \nu} + (1-\xi)\frac{k^\mu k^\nu}{k^2}\right)\;,\\
    \raisebox{-0.25cm}{\includegraphics[scale=1.1]{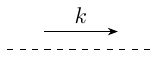}}
    \ & = \ \frac{i}{k^2 - m_\phi^2+ i\epsilon}\;,\\
    \raisebox{-0.25cm}{\includegraphics[scale=1.1]{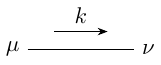}}
    \ & = \ \frac{i}{k^2-m_{\mbox{\tiny V}}^2+ i\epsilon}\left(-\eta^{\mu \nu} + \frac{k^\mu k^\nu}{m_{\mbox{\tiny V}}^2}\right)\;,
\end{aligned}
\label{eq: Phi3FeynmanRules}
\end{equation}
where we choose $\xi = 1$ (Feynman gauge). The vertex rules for scalar-photon
interactions read 
{\allowdisplaybreaks
\begin{align}
    & \raisebox{-1.25cm}{\includegraphics{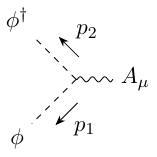}} \ = \ ie(p_1 - p_2)_\mu\;, \\
    & \raisebox{-1.25cm}{\includegraphics{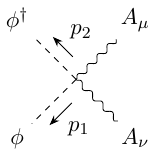}} \ = \ 2i e^2 \eta_{\mu \nu}\;.
\end{align}
}
and the vertex rules for vector-photon interactions read 
\begin{align}
    & \raisebox{-1.25cm}{\includegraphics{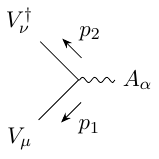}} \ = \ -ie\left[\eta_{\mu \nu}(p_1 - p_2)_\alpha - \eta_{\mu \alpha}(2p_1 + p_2)_\nu + \eta_{\nu \alpha}(2p_2 + p_1)_\mu\right]\;, \\
    & \raisebox{-1.25cm}{\includegraphics{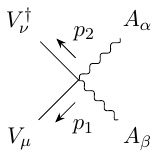}} \ = \ ie^2 \left[\eta_{\mu \alpha}\eta_{\nu \beta} + \eta_{\mu \beta}\eta_{\nu \alpha} - 2\eta_{\mu \nu}\eta_{\alpha \beta}\right]\;.
\end{align}

\section{Covariant Lorentz transformation for polarization vectors}
\label{appendix:CovariantBoostedpolarizations}
Here we discuss the derivation of the Lorentz transformations used in 
\equationref{eq:polarizationBoost}. 
In order to do this in a covariant way, we want to first determine some general
Lorentz transformation involving the appropriate variables, and then enforce the
following 
\begin{equation}
\begin{split}
    p_1 = -\bar{p}_1 + \frac{q}{2}  \quad &\longrightarrow\qquad \bar{p}_1^{\,\prime} = \frac{m_1}{\bar{m}_1}\bar{p}_1\;, \\ 
    p_4 = \bar{p}_1 + \frac{q}{2} \!\qquad&\longrightarrow\qquad \bar{p}_1^{\,\prime} = \frac{m_1}{\bar{m}_1}\bar{p}_1\;,
\end{split}
\end{equation}
where the normalization $(m_1/\bar{m}_1)$ ensures that the magnitude of momentum 
is unchanged. It is then natural to define our infinitesimal generator as
\begin{equation}
  \omega^{\mu\nu} = \beta \left(\bar{p}_1^\mu q^\nu - q^\mu \bar{p}_1^{\nu}\right)\;,
\end{equation}
where $\beta$ is a free parameter still to be defined. We then write the
exact transformation via exponentiation
\begin{equation}
  \Lambda^{\mu}_{~\nu} = \left(e^{-\frac{i}{2}\omega_{\rho \sigma}\Sigma^{\rho \sigma}}\right)^\mu_{~\nu} 
  = \left(e^{\beta X}\right)^\mu_{~\nu} = \sum_{n = 0}^{\infty}{\frac{(\beta^n X^n)^\mu_{~\nu}}{n!}}\;,
\end{equation}
where the infinitesimal Lorentz generator is
\begin{equation}
	(\Sigma^{\rho \sigma})^{\mu}_{~\nu} = i(\eta^{\rho \mu}\delta^{\sigma}_{~\nu}-\eta^{\sigma \mu}\delta^{\rho}_{~\nu})\;,
\end{equation}
and generally we have
\begin{equation}
   (X^{2n+1})^\mu_{~\nu} = (-q^2 \bar{m}_1^2)^n X^\mu_{~\nu}\;, \qquad 
   (X^{2n})^\mu_{~\nu} = (-q^2 \bar{m}_1^2)^n(X^2)^\mu_{~\nu}\;,
\end{equation}
together with
\begin{equation}
  X^\mu_{~\nu} = \bar{p}_1^\mu q_\nu - q^\mu \bar{p}_{1\nu}\;, \qquad 
  (X^2)^\mu_{~\nu} = -q^2\bar{p}_1^\mu \bar{p}_{1\nu} - \bar{m}_1^2q^\mu q_\nu\;.
\end{equation}
Splitting the summation into even and odd powers, one can write the transformation
as
\begin{equation}
\begin{aligned}
  \Lambda(\psi)^{\mu}_{~\nu} & = \delta^\mu_{~\nu} + Y^\mu_{~\nu}\sum_{n = 0}^{\infty} \frac{(\psi)^{2n+1}}{(2n+1)!} 
  + (Y^2)^\mu_{~\nu}\sum_{n = 0}^{\infty} \frac{(\psi)^{2n}}{(2n)!} - (Y^2)^\mu_{~\nu} \\
   & \begin{aligned}
     & = \delta^\mu_{~\nu} + \sinh(\psi)Y^\mu_{~\nu} + (\cosh(\psi)-1)(Y^2)^\mu_{~\nu}\;,
    \end{aligned}
\end{aligned}
\end{equation}
where $\psi = \beta\sqrt{-q^2 \bar{m}^2_1}$ and $Y^\mu_{~\nu} =
\frac{1}{\sqrt{-q^2 \bar{m}^2_1}} X^\mu_{~\nu}$. Finally, to find the
appropriate angles for our transformations, we enforce the following equalities
\begin{equation}
   \Lambda(\psi)^{\mu}_{~\nu} \bar{p}_1^{\,\prime\nu} \stackrel{!}{=} -\bar{p}^\mu_1 + \frac{q^\mu}{2}\;, \qquad 
   \Lambda(\psi')^{\mu}_{~\nu}  \bar{p}_1^{\,\prime\nu} \stackrel{!}{=} \bar{p}^\mu_1 + \frac{q^\mu}{2}\;,
\end{equation}
where $\psi'$ is the angle needed for the transformation of $p_4$.
To illustrate this explicitly, let us consider solving one of the equations
above. For example, let us consider 
\begin{equation}
\begin{split}
    -\bar{p}^\mu_1 + \frac{q^\mu}{2} & \stackrel{!}{=} \Lambda(\psi)^{\mu}_{~\nu}  \bar{p}_1^{\,\prime\nu} \\
    & = \delta^\mu_{~\nu}\bar{p}_1^{\,\prime\nu} + \sinh(\psi)Y^\mu_{~\nu}\bar{p}_1^{\,\prime\nu} + (\cosh(\psi)-1)(Y^2)^\mu_{~\nu}\bar{p}_1^{\,\prime\nu} \\
    & = -\frac{m_1}{\sqrt{-q^2}}\sinh(\psi)q^\mu + \cosh(\psi)\bar{p}_1^{\,\prime\mu}\;,
\end{split}
\end{equation}
which gives two equations
\begin{equation}
    -\frac{m_1}{\sqrt{-q^2}}\sinh(\psi) - \frac{1}{2} = 0\;, \qquad \frac{m_1}{\bar{m}_1}\cosh(\psi) + 1 = 0\;, 
\end{equation}
whose solutions are
\begin{equation}
    \sinh(\psi) = -\frac{\sqrt{-q^2}}{2 m_1}\;, \qquad \cosh(\psi) = - \frac{\bar{m}_1}{m_1}\;.
\end{equation}
Then, putting these solutions together gives us
\begin{equation}
    \psi = \arctanh\left(\frac{\sqrt{-q^2}}{2\bar{m}_1}\right),
\end{equation}
which is the angle needed for the transformation of $p_1$. The angle $\psi'$ for 
the transformation of $p_4$ to $\bar{p}_1^{\,\prime}$ is simply given by $\psi' = -\psi$.


\section{Evaluating tensor integrals in Fourier transforms}
\label{appendix:EvaluatingRankedIntegrals}
Here we discuss the computation of the tensor integrals relevant for the
calculation of observables as done in \sectionref{sec:Observables}. The important point is that each integral is
a function of $b^2$, meaning that we can determine higher rank integrals from
the lower rank integrals via differentiation. As such, the first step is to
compute the scalar integral
\begin{equation}
    f_{\beta} (b^2) = \int \hat{d}^D q \ \hat{\delta}(2\bar{m}_1 \bar{u}_1 \cdot q) \hat{\delta}(-2\bar{m}_2 \bar{u}_2 \cdot q)e^{i b \cdot q} (-q^2)^{-\beta}\;.
\end{equation}
which can be done using a Sudakov decomposition of the $D$-dimensional exchange momentum
\begin{equation}
    q^\mu = x_1 \bar{u}_1^\mu + x_2 \bar{u}_2^\mu + \boldsymbol{q}_\perp^\mu\;,
\end{equation}
where $\boldsymbol{q}_\perp^\mu$ is in the $(D-2)$-dimensional subspace
transverse to $\bar{u}_1$ and $\bar{u}_2$, e.g.\ in the decomposition
of Ref.~\cite{Herrmann:2021tct}. Then, this
parametrization yields
\begin{equation}
\begin{split}
    f_{\beta} (b^2) & = \frac{1}{N} \int \hat{d}^{D-2} \boldsymbol{q}_\perp\hat{d}x_1\hat{d}x_2\ \hat{\delta}(x_1) \hat{\delta}(x_2) e^{i b \cdot q} (-q^2)^{-\beta}\\
    & = \frac{1}{N} \int \hat{d}^{D-2} \boldsymbol{q}_\perp \ e^{-i \boldsymbol{b} \cdot \boldsymbol{q}_\perp} (\boldsymbol{q}_\perp^2)^{-\beta} \\
    & = \frac{4}{N} \frac{\Gamma(D/2 - 1 - \beta)}{2^{2\beta + 2}\pi^{(D - 2)/2}\Gamma(\beta)} \frac{1}{|b|^{D-2-2\beta}}\;,
\end{split}
\end{equation}
where $N = 4\bar{m}_1 \bar{m}_2\sqrt{y^2 - 1}$.
Here, the delta functions localize the two integration variables and force the
momentum transfer to span the $(D-2)$-dimensional subspace. The impact
parameter is purely transverse with $\boldsymbol{b}^2 = -b^2 \equiv |b|^2$.
The final integration is done by orienting the vectors appropriately and
using spherical-polar coordinates. 

Then, the higher-rank integrals follow trivially. For instance, for rank-$1$ we
find
\begin{equation}
f^\mu_{\beta} (b^2)  = \frac{1}{i} \frac{\partial f_{\beta} (b^2)}{\partial b_\mu}  
 = -\frac{4i}{N} \frac{\Gamma(D/2 - \beta)}{2^{2\beta + 1}\pi^{(D - 2)/2}\Gamma(\beta)} \frac{b^\mu}{|b|^{D-2\beta}}\;,
\end{equation}
while for rank-$2$ tensor integrals we obtain
\begin{equation}
\begin{split}
f^{\mu \nu}_{\beta} (b^2) & = \frac{1}{i} \frac{\partial f^\mu_{\beta} (b^2)}{\partial b_\nu} \\
& = -\frac{4}{N} \frac{\Gamma(D/2 - \beta)}{2^{2\beta + 1}\pi^{(D - 2)/2}\Gamma(\beta)} \frac{1}{|b|^{D+2-2\beta}}
\Bigg(|b|^2 \Pi^{\mu \nu} + (D-2\beta) b^\mu b^\nu \Bigg)\;,
\end{split}
\end{equation}
and finally for rank-$3$ integrals we have
\begin{equation}
\begin{split}
f^{\mu \nu \gamma}_{\beta} & (b^2)  = \frac{1}{i} \frac{\partial f^{\mu \nu}_{\beta} (b^2)}{\partial b_\gamma} \\
& = \frac{4i}{N} \frac{\Gamma(D/2 - \beta)}{2^{2\beta + 1}\pi^{(D - 2)/2}\Gamma(\beta)} \frac{(D-2\beta)}{|b|^{D+4-2\beta}}
\Bigg(3 |b|^2 b^{(\mu}\Pi^{\nu \gamma)} + (D+2-2\beta) b^\mu b^\nu b^\gamma\Bigg)\;.
\end{split}
\end{equation}
In these expressions we have replaced all instances of the metric with the
transverse projector
\begin{equation}
\begin{split}
 \Pi^{\mu}_{~\nu} & = \frac{1}{y^2 - 1}\epsilon^{\mu \rho \alpha \beta} \epsilon_{\nu \rho \gamma \delta} \bar{u}_{1\alpha }\bar{u}_{2\beta}\bar{u}_{1}^{\gamma} \bar{u}_{2}^{\delta} \\
    & =  \delta^{\mu}_{~\nu} + \frac{1}{y^2 - 1}\Big(\bar{u}_1^\mu (\bar{u}_{1\nu} - y \bar{u}_{2\nu}) + \bar{u}_2^\mu (\bar{u}_{2\nu} - y \bar{u}_{1\nu})\Big)\;,
\end{split}
\end{equation}
which takes us to a plane orthogonal to the velocities, i.e.  the transverse
plane.


\bibliographystyle{JHEP}
\bibliography{casimir}
\end{document}